\documentclass[preprint]{ptephy_v1}

\preprintnumber{YITP-16-85} 

\usepackage{color}
\usepackage{longtable}
\usepackage{graphicx}
\usepackage{dcolumn}
\usepackage{bm}
\usepackage{amsmath,amssymb}

\usepackage{cancel}
\usepackage{ulem} 

\newcommand{\re}{\text{Re }}
\newcommand{\im}{\text{Im }}
\newcommand{\physdim}[1]{\hspace{1ex} \mathrm{#1}}
\usepackage{wrapfig} %

\begin{document}

\title{Generalized weak-binding relations of compositeness in effective field theory}


\author{Yuki~Kamiya}
\affil{Yukawa Institute for Theoretical Physics, Kyoto University, Kyoto 606-8502, Japan \email{yuki.kamiya@yukawa.kyoto-u.ac.jp}}


\author[2]{Tetsuo~Hyodo}
\affil{Yukawa Institute for Theoretical Physics, Kyoto University, Kyoto 606-8502, Japan}


\begin{abstract}
We study the compositeness of near-threshold states to investigate the internal structure of exotic hadron candidates. 
Within the framework of effective field theory, Weinberg's weak-binding relation is extended to more general cases by easing several preconditions.
First, by evaluating the contribution from the decay channel, we obtain the generalized relation for unstable quasibound states. 
Next, we generalize the relation to include the nearby CDD (Castillejo--Dalitz--Dyson) pole contribution with the help of the Pad\'e approximant. 
The validity of the estimation with the generalized weak-binding relations is examined by numerical calculations.
A method to systematically evaluate the error in the weak-binding relation is presented.
Finally, by applying the extended relation to $\Lambda(1405)$, $f_0(980)$, and $a_0(980)$, we discuss their internal structure, in comparison with other approaches.
\end{abstract}

\subjectindex{D32}
\maketitle

\section{Introduction}

The investigation of the internal structure of hadrons is one of the most fundamental subjects in hadron physics.
The discovery of many candidates for exotic hadrons, which are not assigned to simple $q\bar{q}$ mesons or $qqq$ baryons, drives us to study their internal structure. 
One of the famous candidates for such hadrons is the $\Lambda(1405)$ resonance, whose mass is much lighter than the other negative parity baryons~\cite{Agashe:2014kda}.
Recently it has been considered that the $\bar{K}N$ composite component is essential for the nature of $\Lambda(1405)$ resonance~\cite{Kaiser:1995eg,Oset:2001cn,Oller:2000fj,Hall:2014uca,Miyahara:2015uya,Kamiya:2015aea,Kamiya:2016jqc}.
Moreover many $XYZ$ mesons, which are not understood as simple $c\bar{c}$ or $b\bar{b}$  states, are observed in the heavy quark sector~\cite{Swanson:2006ap,Brambilla:2010cs,Hosaka:2016pey}.
Specifying the structure of these hadrons, we can acquire knowledge of strong interaction in the hadrons.   

To investigate the hadron structure, a number of models have been constructed by reproducing the experimental data. 
However, there is an ambiguity in identifying the structure in model-dependent studies.
When we construct a model to reproduce the experimental data, the contribution of the degrees of freedom that are excluded from the model space is taken into account by the model parameters. 
It is not clear whether the employed degrees of freedom are suitable, even when the model reproduces the experimental data well~\cite{Hyodo:2008xr,Hyodo:2013nka}.

In contrast, the internal structure is determined without ambiguity in model-independent approaches in the weak-binding limit.
One such method discusses the compositeness of the state with the weak-binding relation, which is derived by  Weinberg~\cite{Weinberg:1965zz}. 
This relation implies that, neglecting the correction terms, the compositeness of a weakly bound state is directly determined by experimental observables, the scattering length and the eigenenergy.
Using the weak-binding relation, it is shown that deuteron is a composite system of a proton and neutron without any assumption of the nuclear force potential~\cite{Weinberg:1965zz}.
Hence, this is a suitable technique for studying the structure of hadrons.

To apply this method to the candidates for exotic hadrons, we have to extend the weak-binding relation to be more practical for various cases. 
First, because most of the exotic hadrons are unstable and have a finite decay width, we need a relation valid for an unstable state.
Several approaches in this direction have been proposed.
In Ref.~\cite{Baru:2003qq}, they extend Weinberg's discussion by introducing the spectral function and show that this function includes a measure of the admixture of the bare state in the physical state.
The multichannel case is discussed in Ref.~\cite{Hanhart:2011jz}. 
In Ref.~\cite{Hyodo:2013iga}, Weinberg's weak-binding formula is directly applied to a resonance close to the lowest energy threshold in the convergence region of the effective range expansion.
On the other hand, a method to evaluate the compositeness of hadron resonances away from the threshold has also been developed.
In Ref.~\cite{Hyodo:2011qc}, it is shown that the compositeness of $s$-wave states is expressed by the residue of the pole and the derivative of the loop function based on the Yukawa model. 
A similar expression is obtained in Ref.~\cite{Aceti:2012dd} for general partial waves using the wave function in the cutoff model.
It is shown in Ref.~\cite{Sekihara:2014kya} that the compositeness of a resonance can be rigorously defined from the wave function with general separable potentials. 
In Ref.~\cite{Guo:2015daa}, they show that a rank 1 projection operator can be extracted from the residues of the pole of a resonance and suggest the quantity that can be interpreted as a probability of compositeness.  

Second, as already mentioned in Ref.~\cite{Weinberg:1965zz}, the weak-binding relation is not valid when the CDD (Castillejo-Dalitz-Dyson) pole~\cite{Castillejo:1955ed} lies near the threshold. The CDD pole is defined as the pole of the inverse amplitude. It is considered that the existence of the CDD pole reflects the contribution that comes from outside the model space~\cite{Chew:1961}.
Recently, the relation between the CDD pole position and the compositeness has been discussed in detail in Ref.~\cite{Baru:2010ww}.
It is pointed out that the contribution of the CDD pole nearby the $\pi\Sigma_c$ threshold energy may become important for the $\Lambda_c(2595)$ resonance~\cite{Guo:2016wpy}. 

Third, the weak-binding relations always contain higher-order correction terms, as we will see below. Although the terms are small for the near-threshold states, they cause uncertainty in the results of the weak-binding relations. For practical applications to the hadron structures, it is useful to develop a method to estimate the uncertainties of the weak-binding relations that arise from the correction terms.

In this paper, we directly extend the weak-binding relation to a near-threshold quasibound state with a lower-energy decay channel and to a state with a nearby CDD pole based on the effective field theory.
In Sects.~\ref{sec:bound} and \ref{sec:quasibound}, we first present a detailed derivation of the results in Ref.~\cite{Kamiya:2015aea}, where the weak-binding relation is extended to the quasibound state by using nonrelativistic effective field theory~\cite{Kaplan:1996nv,Braaten:2007nq}.
We newly construct a method to estimate the error of the compositeness in the weak-binding relations, both for stable and unstable states.
Next, in Sect.~\ref{sec:CDD}, we consider the case that the effective range expansion of the scattering amplitude does not converge well. 
The weak-binding relation is then extended to this case by carefully considering two independent expansions in the derivation.
In Sect.~\ref{sec:model}, we discuss the validity of the estimation of the compositeness with original and extended weak-binding relations in numerical model calculations. 
The model calculations are also used to examine method of error evaluation developed in Sect.~\ref{sec:bound}.
From these results, we discuss the determination of the compositeness by the weak-binding relation, in comparison with model calculations.
Finally, as applications to the hadrons, we use the extended weak-binding relations to discuss the structure of $\Lambda(1405)$, $f_0(980)$ and $a_0(980)$.
Conclusions are given in the final section.

\section{Weak-binding relation for bound state}\label{sec:bound}

\subsection{Effective field theory}
First, we review the derivation of the weak-binding relation written in Ref.~\cite{Kamiya:2015aea}.
Here we consider the structure of a stable bound state that appears in the $s$-wave scattering of two hadrons.
We assume that the typical range scale of the interaction is $R_{\mathrm{typ}}$ and there is no long range force.
We use the nonrelativistic effective field theory (EFT)~\cite{Kaplan:1996nv,Braaten:2007nq} to analyze the property of this system in the low energy region.
We introduce the fields $\psi$, $\phi$, $B_0$ whose quantum numbers correspond to those of the hadrons in the scattering channel and the bound state, respectively.
The Hamiltonian of the nonrelativistic effective field theory for the low energy domain of this system is given as 
\begin{align}
H &=H_{\mathrm{free}} + H_{\mathrm{int}} =  \int d^3\bm{x} (\mathcal{H}_{\mathrm{free}}+\mathcal{H}_{\mathrm{int}})\label{eq:Hamiltonian-bound}, \\
\mathcal{H}_{\mathrm
{free}} &=\frac{1}{2 M} \mathbf{\nabla} \psi^\dagger(\bm{x}) \cdot\mathbf{\nabla} \psi(\bm{x}) +\frac{1}{2 m} \mathbf{\nabla} \phi^\dagger(\bm{x}) \cdot\mathbf{\nabla} \phi (\bm{x})\notag\\
&\quad+ \frac{1}{2M_{0}} \mathbf{\nabla}  B_0^\dagger(\bm{x}) \cdot{\mathbf \nabla} B_0 (\bm{x})+ 
   \omega_0 B_0^\dagger(\bm{x}) B_0 (\bm{x}),\\
\mathcal{H}_{\mathrm{int}} &= g_0\left( B_0^\dagger(\bm{x}) \psi(\bm{x})\phi (\bm{x})+ \phi^\dagger(\bm{x})\psi^\dagger (\bm{x})B_0(\bm{x}) \right)  +  v_{0} \psi^\dagger(\bm{x})\phi^\dagger (\bm{x})\phi(\bm{x})\psi(\bm{x}),\label{eq:H_int}
\end{align}
where $\hbar =1$.
These fields are quantized by the appropriate commutation relations, 
$
[ \psi(\bm{x}),\psi^\dagger(\bm{x}^\prime)]_{\pm}=[ \phi(\bm{x}),\phi^\dagger(\bm{x}^\prime)]_{\pm}=
[ B_0(\bm{x}),B_0^\dagger(\bm{x}^\prime)]_{\pm}
=\delta^3(\bm{x} - \bm{x}^\prime)
 $.
In the low energy region, the details of the short range interaction are not relevant and the system can be described by the contact interaction of Eq.~\eqref{eq:H_int}. 
To tame the ultraviolet divergence of the momentum integration, we use a sharp cutoff $\Lambda$, which 
corresponds to the upper limit of the momentum where the description of the system by the contact interaction is appropriate.
Thus the cutoff $\Lambda$ is related to the typical length scale of interaction $R_{\mathrm{typ}}$ as $\Lambda  \sim 1/R_{\mathrm{typ}}$.

Using Noether's theorem for the phase symmetry of the corresponding Lagrangian, we obtain the following conservation laws of particle numbers:
\begin{align}
N_{\mathrm{I}} &\equiv N_{\psi}+ N_{B_0} =\mathrm{const.}\label{eq:N1-bound}\\
N_{\mathrm{II}} &\equiv N_{\phi} + N_{B_0}=\mathrm{const.} \label{eq:N2-bound}
\end{align}
Here $N_\alpha = \int d^3\bm{x} \alpha^\dagger(\bm{x}) \alpha(\bm{x})$ denotes the number of particle $\alpha$.
Now we consider the sector with $(N_\mathrm{I},N_{\mathrm{II}}) = (1,1)$ to discuss the scattering of $\psi$ and $\phi$.
There are only two states with $(N_\mathrm{I},N_{\mathrm{II}}) = (1,1)$, $\{N_\psi=1, N_\phi=1, N_{B_0}=0\}$ and $\{N_\psi=0,N_\phi=0,N_{B_0}=1\}$.
The corresponding eigenstates of $H_{\mathrm{free}}$ are the scattering state  $|\bm{p}\rangle$ and the discrete state $|B_0\rangle$ defined as
\begin{align}
|\bm{p}\rangle  \equiv \frac{1}{\sqrt{ V_p}}\tilde{\psi}^\dagger(\bm{p}) \tilde{\phi}^\dagger(-\bm{p})|0\rangle,\quad
|B_0\rangle\equiv \frac{1}{\sqrt{V_p}}\tilde{B}_0^\dagger(\bm{0})|0\rangle,
\end{align}
where the vacuum is defined as
\begin{align}
\tilde{\psi}(\bm{p})|0\rangle = \tilde{\phi}(\bm{p})|0\rangle = \tilde{B}_0(\bm{p})|0\rangle=0 \hspace{3ex} (\mathrm{for}\hspace{1ex} \forall \bm{p}) ,\hspace{3ex}\langle 0 | 0 \rangle = 1,\label{eq:vacuum}
\end{align}
with the annihilation operator $\tilde{\alpha}(\bm{p})=\int d^3\bm{x} \exp(-i\bm{p}\cdot\bm{x})\alpha(\bm{x})$ and the phase space of the system $V_p = (2\pi)^3 \delta^3(\bm{0})$.
Here we take the reference frame in the center of mass system of the scattering state.
These eigenstates are normalized as $ \langle \bm{p} | \bm{p^\prime} \rangle =(2\pi)^3\delta^3(\bm{p} - \bm{p}^\prime), \langle B_0|B_0\rangle = 1$, and are orthogonal to each other.
Eigenenergies are given by
\begin{align}
H_{\mathrm{free}} |\bm{p} \rangle=  \frac{\bm{p}^2}{2\mu}|\bm{p}\rangle,\quad  H_{\mathrm{free}} |B_0\rangle = \omega_0|B_0 \rangle.
\end{align}
From particle number conservation, the completeness relation of this sector is written as
\begin{align}
	\int \frac{d^3\bm{p}}{(2\pi)^3}  |\bm{p}\rangle\langle\bm{p}| + |B_0\rangle\langle B_0| =1\label{eq:completeness}.
\end{align}
This is a model space equivalent to that used in Refs.~\cite{Weinberg:1965zz,Sekihara:2014kya}. In the present EFT formulation, the definition of the discrete state $|B_0\rangle$ is clear and the completeness relation Eq.~\eqref{eq:completeness} follows from particle number conservation law. 
We also note that the contact interaction is of separable type . In contrast to Ref.~\cite{Sekihara:2014kya} where the separable nature of the interaction is an assumption, in the present study, the contact interaction naturally arises to describe the low energy $s$-wave interaction.

\subsection{Scattering amplitude}

Next we construct the scattering amplitude of $\psi$ and $\phi$.
For this purpose, we derive the effective Hamiltonian with the discrete state being eliminated.
We first write the Schr\"odinger equation for the bound state as 
\begin{align}
H|\Psi\rangle = E_h|\Psi\rangle \label{eq:Schrodinger},
\end{align}
where $E_h <0 $ is the eigenenergy.
Because the completeness relation of this sector is written as Eq. \eqref{eq:completeness}, we can write $|\Psi \rangle$ as the superposition of $|\bm{p}\rangle$ and $|B_0\rangle$:
\begin{align}
	|\Psi\rangle = \int \frac{d^3\bm{p}}{(2\pi)^3}\chi(\bm{p})|\bm{p}\rangle +c |B_0\rangle \label{eq:def-Psi},
\end{align}
where $\chi(\bm{p})=\langle \bm{p} |\Psi\rangle$ and $c=\langle B_0|\Psi\rangle$ denote the weights of the scattering state and discrete state in the bound state $|\Psi\rangle$, respectively.

By using the Schr\"odinger equation \eqref{eq:Schrodinger} for $\langle \bm{p} | H | \Psi\rangle$ and  $\langle B_0 |H|\Psi \rangle$, the following relations between $\chi(\bm{p})$ and $c$ are derived:
\begin{align}
	\chi(\bm{p})E_h 
	 &= \frac{p^2}{2\mu}\chi(\bm{p}) +v_0\int \frac{d^3\bm{p}^\prime}{(2\pi)^3} \chi(\bm{p}^\prime) + cg_0 \label{eq:mt-element1},\\
	cE_h  
	 &=\int \frac{d^3\bm{p}^\prime}{(2\pi)^3} \chi(\bm{p}^\prime) g_0 + c\omega_0.\label{eq:mt-element2}
\end{align}
Eliminating $c$ from these relations, we obtain following equation for $\chi(\bm{p})$: 
\begin{align}
&	\frac{p^2}{2\mu} \chi(\bm{p}) + v(E_h) \int \frac{d^3\bm{p}^\prime}{(2\pi)^3} \chi(\bm{p}^\prime) = E_h \chi(\bm{p})\label{eq:chi-Schrodinger}, \\
&	v(E) =  v_0 + \frac{g_0^2}{E - \omega_0} \label{eq:v-eff-bound}.
\end{align}
Here $v(E)$ denotes the effective interaction for the scattering state $|\bm{p}\rangle$ which includes the contribution of the transition to the discrete state $|B_0\rangle$.
Now we introduce the effective Hamiltonian as
\begin{align}
H_{\mathrm{eff}}(E)&=H_0+V(E) \label{eq:H_eff}\\
H_0 &= \int \frac{d^3\bm{p}}{(2\pi)^3}\left[\frac{\bm{p}^2}{2M}\tilde{\psi}^\dagger(\bm{p})\tilde{\psi}(\bm{p}) + \frac{\bm{p}^2}{2m}\tilde{\phi}^\dagger(\bm{p})\tilde{\phi}(\bm{p})\right]\label{eq:H_0},\\
V(E) &= \int \prod_{i=1}^{4}\frac{d^3\bm{p}_i}{(2\pi)^3}(2\pi)^3\delta^3(\bm{p}_1 +\bm{p}_2 - \bm{p}_3- \bm{p}_4) v(E)\tilde{\psi}^\dagger(\bm{p}_1)\tilde{\phi}^\dagger(\bm{p}_2)\tilde{\phi}(\bm{p}_3)\tilde{\psi}(\bm{p}_4)\label{eq:V}.
\end{align}
For this Hamiltonian, the Schr\"odinger equation is written as
\begin{align}
H_{\mathrm{eff}}(E_h)|\Psi_{\mathrm{B}}\rangle =& E_h|\Psi_{\mathrm{B}}\rangle\label{eq:Shrodinger-chi},\\
|\Psi_{\mathrm{B}}\rangle=& \int\frac{d^3\bm{p}}{(2\pi)^3}\chi(\bm{p})|\bm{p}\rangle.  \label{eq:exp-eigenstate}
\end{align}
Expanding the eigenstate as in Eq.~\eqref{eq:exp-eigenstate}, we can derive Eq.~\eqref{eq:chi-Schrodinger} from Eq.~\eqref{eq:Shrodinger-chi}.
The eigenenergy and the wave function $\chi(\bm{p})$ of this new Schr\"odinger equation \eqref{eq:Shrodinger-chi} are equivalent to the original ones.

Next we derive the scattering amplitude using Eq.~\eqref{eq:H_eff}.
The $T$ operator $T(E)$ satisfies the following Lippmann--Schwinger equation:
\begin{align}
T(E)= V(E) +V(E)\frac{1}{E-H_{0} +i0^+}T(E).\label{eq:Rippmann}
\end{align}
The effective interaction $V(E)$ in Eq.~\eqref{eq:V} leads to $\langle\bm{p}^\prime|V(E)|\bm{p}\rangle=v(E)$, which is independent of the initial and final momenta.
Thus the matrix element $T(E;\bm{p};\bm{p}^\prime)= \langle \bm{p}|T(E)|\bm{p}^\prime\rangle$ is a function of the energy $E$. 
The Lippmann--Schwinger equation of the on-shell $T$ matrix $t(E)= T(E;\bm{p};\bm{p}^\prime)|_{E=p^2/(2\mu)=p^{\prime2}/(2\mu)}$ is then given by 
\begin{align}
t(E)&=v(E) + v(E)G(E)t(E), \\
G(E)&\equiv \int\frac{d^3\bm{p}}{(2\pi)^3}\frac{1}{E-p^2/(2\mu)+i0^+}, 
\end{align}
where $G(E)$ is the loop function whose ultraviolet divergence is tamed by the cutoff $\Lambda$.
This Lippmann--Schwinger equation can be solved algebraically as 
\begin{align}
	t(E) &= \frac{1}{v^{-1}(E) - G(E) } \label{eq:T-mat}.
\end{align}
Because there has been no approximation in the above derivation, Eq.~\eqref{eq:T-mat} is the exact solution of the two-body scattering problem.

With this solution of the $T$ matrix, the scattering amplitude $\mathcal{F}(E)= -(\mu/2\pi)t(E)$ can be written as
\begin{align}
	\mathcal{F}(E) =  -\frac{\mu}{2\pi}\frac{1}{[v(E)]^{-1}-G(E) } \label{eq:amp}.
\end{align}
Because the bound state is expressed as the pole of the scattering amplitude, the condition of the bound state is given by
\begin{align}
	 [v(E_h)]^{-1}- G(E_h)=0	\label{eq:pole-condition}.
\end{align}
We can also write down the scattering length as
\begin{align}
	a_0 =   \frac{\mu}{2\pi}\frac{1}{[v(0)]^{-1} -G(0)} \label{eq:scat-length},
\end{align}
where we define  $a_0 =-\mathcal{F}(0)$.

\subsection{Compositeness}

We normalize the bound state $|\Psi\rangle$ with the following condition:
\begin{align}
\langle \Psi | \Psi\rangle &= 1 \label{eq:normalization}.
\end{align}
The completeness relation \eqref{eq:completeness} indicates that the sum of the projections to the scattering state $|\bm{p}\rangle\langle\bm{p}|$ and to the discrete state $|B_0\rangle\langle B_0|$ equals $1$.
We define the compositeness $X$ (the elementariness $Z$) as the sum of the projections to the scattering state (the projection to discrete state) in the normalization~\eqref{eq:normalization}:
\begin{align}
	X \equiv &\int \frac{d^3\bm{p}}{(2\pi)^3}\langle \Psi |\bm{p}\rangle \langle \bm{p}|\Psi\rangle= \int \frac{d^3\bm{p}}{(2\pi)^3} | \chi(\bm{p}) |^2 \label{eq:def-X},\\
	Z \equiv &\langle \Psi|B_0\rangle\langle   B_0|\Psi\rangle= |c|^2 \label{eq:def-Z} .
\end{align}
With Eq.~\eqref{eq:normalization}, we obtain the sum rule:
\begin{align}
	X +Z = 1 \label{eq:X+Z}.
\end{align}
By definition $X,Z\geq0$, then the values of $X$ and $Z$ satisfy 
\begin{align}
	X,Z \in [0,1] \label{eq:X-Z-in-0-to-1}.
\end{align}
Equations~\eqref{eq:X+Z} and \eqref{eq:X-Z-in-0-to-1} ensure that we can consider $X$ and $Z$ as the probability of finding the scattering states and the discrete state in the bound state, respectively.

Using Eqs.~\eqref{eq:mt-element2}, \eqref{eq:chi-Schrodinger}, and \eqref{eq:v-eff-bound} to determine $\chi(\bm{p})$ and $c$, then $X$ and $Z$ can be written as
\begin{gather}
X=-|\alpha|^2 v^2(E_h) G^\prime(E_h),\quad Z=-|\alpha|^2v^{\prime}(E_h)\label{eq:X-Z-alpha},\quad 
\alpha \equiv \int \frac{d^3\bm{p}}{(2\pi)^3} \chi(\bm{p}),
\end{gather}
where $A^\prime(E)$ denotes the derivative of the function $A(E)$ with respect to $E$.
Eliminating $\alpha$ by Eq.~\eqref{eq:X+Z}, we obtain
\begin{align}
	X = \frac{G^\prime(E_h)}{G^\prime(E_h) - \left[1/v(E_h) \right]^\prime},\quad Z = \frac{-\left[1/v(E_h) \right]^\prime}{G^\prime(E_h)-\left[1/v(E_h) \right]^\prime   }\label{eq:X-Z-bound}.
\end{align}
Let us consider the cutoff dependence of the compositeness.
The effective field theory universally describes the physics in the low energy region, and the contribution from the short range interaction of the microscopic theory is represented by the cutoff scale. 
The quantities that cannot be observed by experiments, such as the wave function $\chi(\bm{p})$, can depend on the cutoff.
Hence the compositeness determined by the wave function in Eq.~\eqref{eq:def-X} is in general a cutoff-dependent quantity.

Now we discuss the role of the discrete state $B_0$ in the EFT. 
When the scattering channel $\psi\phi$ does not couple to the discrete state ($g_0=0$), the effective interaction $v(E)$ has no energy dependence.
In this case $[1/v(E_h)]^\prime$ in Eq.~\eqref{eq:X-Z-bound} becomes 0 and thus $X=1$ definitely.
If we introduce more general effective theories with two or more discrete states, the channels other than the scattering state contribute to the energy-dependent term of the effective interaction $V(E)$.
In our effective field theory Eq.~\eqref{eq:Hamiltonian-bound}, these contributions are represented by the discrete state $B_0$. 

\subsection{Weak-binding relation}
Next assuming the binding energy of the eigenstate $-E_h$ is small, we derive the weak-binding relation for the scattering length and compositeness ~\cite{Weinberg:1965zz,Sekihara:2014kya}.
We define the radius $R$ of the wave function as
\begin{align}
	R &\equiv \frac{1}{\sqrt{-2\mu E_h}} \label{eq:def-R} .
\end{align}
A weak-binding condition corresponds to $R\gg R_{\mathrm{typ}}$ with the interaction range $R_{\mathrm{typ}}$.
Using Eqs.~\eqref{eq:X-Z-bound} and \eqref{eq:pole-condition} to expand the scattering length $a_0$ \eqref{eq:scat-length} around $E=E_h$, we find 
\begin{align}
a_0 
&= \frac{\mu}{2\pi} \left[\frac{E_hG^{\prime}(E_h)}{X} +\Delta v^{-1} - \Delta G\right]^{-1}\label{eq:a0-bound},\\
\Delta v^{-1}  &\equiv \sum_{n=2} \frac{1}{n!}\left. \frac{d^nv^{-1}}{dE^n}\right|_{E=E_h}(-E_h)^n 
,\quad \Delta G \equiv \sum_{n=2} \frac{1}{n!} \left.\frac{d^nG}{dE^n}\right|_{E=E_h}(-E_h)^n \label{eq:Delta-v-G}.
\end{align}
The loop function $G(E)$ is written as 
\begin{align}
G(E) &=\frac{\mu}{\pi^2}\left[-\Lambda +\sqrt{2\mu E} \mathrm{arctanh}\left(\frac{\Lambda}{\sqrt{2\mu E}}\right)\right] \notag \\
& =\frac{\mu}{\pi^2}\left[-\Lambda-\frac{i}{2}\pi\sqrt{2\mu E}+\frac{2\mu E}{\Lambda} +\frac{(2\mu E)^2}{3\Lambda^3}+\cdots \right].\label{eq:loop-fcn}
\end{align}
Using this equation, we obtain
\begin{align}
G^\prime(E_h) &= -\frac{i\mu^{3/2}}{\sqrt{8}\pi}\frac{1}{\sqrt{E_h}} + \frac{2\mu^2}{\pi^2\Lambda} +\frac{8\mu^3E_h}{3\pi^2\Lambda^3} +\cdots \notag\\
&=  \frac{\mu}{4\pi E_hR }\left\{1+ \mathcal{O}\left(\frac{R_{\mathrm{typ}}}{R}\right)\right\},\label{eq:G-bibun-exp}\\ 
\left.\frac{d^nG}{dE^n}\right|_{E=E_h} &=-(2n-3)!!\left(\frac{-1}{2}\right)^{n+1} \frac{i\mu^{3/2}}{\sqrt{2}\pi}E_h^{-(2n-1)/2}\left[1+\mathcal{O}\left(\left( \frac{R_\mathrm{typ}}{R}\right)^{2n-1}\right)\right], \\
\Delta G&= \sum_{n=2}\frac{1}{n!} \left.\frac{d^nG}{dE}\right|_{E=E_h} (-E_h)^n\notag \\
&= \frac{\mu}{2\pi R} \sum_{n=2}\frac{(2n-3)!!}{n!2^n}\left[1+\mathcal{O}\left(\left( \frac{R_\mathrm{typ}}{R}\right)^{2n-1}\right)\right]  \notag \\
&=\frac{\mu}{4\pi R}\left[1+\mathcal{O}\left(\frac{R_{\mathrm{typ}}}{R}\right)\right].\label{eq:Delta-exp}
\end{align} 
We note that the leading term of $\Delta G$ comes from the nonanalytic term $\sqrt{2\mu E}$ in Eq.~\eqref{eq:loop-fcn}.

Now we consider the estimation of $\Delta  v^{-1}$ in Eq.~\eqref{eq:a0-bound}. With the phase shift $\delta(p)$ and the eigenmomentum $p=\sqrt{2\mu E}$, the $s$-wave scattering amplitude $f(p)$ can be generally expressed as 
\begin{align}
f(p) = \frac{1}{p\cot\delta -ip} \label{eq:scat-delta},
\end{align}
where $p\cot\delta$ is an analytic function of $p^2$.
In the present formulation, the term $-ip$ is included in $G(E)$, so the function $v^{-1}(E)$ must be regular in $E=p^2/(2\mu)$. 
In the low energy region, the scattering amplitude is well approximated by the effective range expansion (ERE) as
\begin{align}
f(p) = \frac{1}{-\frac{1}{a_0} -ip + \frac{r_{\mathrm{e}}}{2} p^2 +\mathcal{O}(R_{\mathrm{eff}}^3p^4)} \label{eq:eff-exp},
\end{align}
where $r_{\mathrm{e}}$ is the effective range and $R_{\mathrm{eff}}$ is the length scale characterizing this expansion. 
The expansion $\Delta  v^{-1}$ in Eq.~\eqref{eq:a0-bound} starts from $E_h^2$, which is included in the term $\mathcal{O}(R_{\mathrm{eff}}^3p^4)$ in Eq.~\eqref{eq:eff-exp} at $E=E_h$.

Now we assume the length scale of the ERE is smaller than the interaction range, $R_{\mathrm{eff}}\lesssim R_{\mathrm{typ}}$. In the weak-binding limit $R_{\mathrm{typ}}/R\ll 1$, $R$ must satisfy $(R_{\mathrm{eff}}/R)^3 \ll 1$.
This assumption indicates that $\mathcal{O}(R_{\mathrm{eff}}^3p^4)$ is negligible and the bound state pole position is in the convergence region of the ERE.
Because the order of $\Delta v^{-1}$ is estimated as $\mathcal{O}(R_{\mathrm{eff}}^3p^4)$, in the present case we obtain 
\begin{align}
\Delta  v^{-1} = \frac{\mu}{R}\mathcal{O} \left( \left(\frac{R_{\mathrm{typ}}}{R}\right)^3\right)\label{eq:estimation-v}.
\end{align}

Using Eqs.~\eqref{eq:G-bibun-exp}, \eqref{eq:Delta-exp}, and \eqref{eq:estimation-v}, the scattering length~\eqref{eq:a0-bound} can be estimated as
\begin{align}
a_{0} =  \frac{\mu }{2\pi} \left\{ \frac{\mu}{4\pi XR } + \frac{\mu}{4\pi R } +\frac{\mu}{R} {\mathcal O}\left(\frac{R_{\mathrm{typ}}}{R}\right)  \right\}^{-1} 
= R\left\{ \frac{2X}{1+X} + {\mathcal O}\left(\frac{R_{\mathrm{typ}}}{R}\right)\right\}  \label{eq:comp-rel-bound}.
\end{align}
As seen from the first term of the right-hand side, the leading term of the expansion of the scattering length in $R_{\mathrm{typ}}/R$ is determined only by the compositeness $X$.
The second term represents the higher-order corrections and explicitly depends on the cutoff $\Lambda\sim 1/R_{\mathrm{typ}}$.
Neglecting the correction terms, the compositeness $X$ can be calculated from only the experimental observables $a_0$ and $E_h$. 

Here we note that the compositeness $X$ can be determined from observables, which are independent of the cutoff, only in the weak-binding limit.
In general, however, the compositeness in Eq.~\eqref{eq:X-Z-bound} depends on the cutoff.
In the expression of Eq.~\eqref{eq:comp-rel-bound}, this cutoff dependence of $X$ is canceled out by that of the correction term $\mathcal{O}(R_{\mathrm{typ}}/R)$ and the ratio of the observables $a_0/R$ remains cutoff independent.
When the higher-order term $\mathcal{O}(R_{\mathrm{typ}}/R)$ is small and can be neglected, the cutoff dependence of the compositeness can also be neglected. In this case we determine $X$ from only the experimental observables.

Up to this point, we have derived the weak-binding relation of Ref.~\cite{Weinberg:1965zz} using the nonrelativistic EFT. In the next section, we extend this relation to the unstable quasibound state. In Sect.~\ref{sec:CDD}, we consider again the assumption on the ERE for Eq.~\eqref{eq:estimation-v} in detail and derive the extended weak-binding relation that is valid even when the ERE does not work well.

\subsection{Error evaluation of compositeness}\label{eq:errorbound}
In the actual applications to near-threshold hadrons, the magnitude of the higher-order term $\mathcal{O}(R_{\mathrm{typ}}/R)$ can be small but finite. Here we develop a systematic method to estimate the uncertainty of the compositeness that comes from the higher-order term.

With the weak-binding relation~\eqref{eq:comp-rel-bound}, the compositeness $X$ can be written as
\begin{align}
X = \frac{a_0/R + \mathcal{O}(R_{\mathrm{typ}}/R)}{2-a_0/R-\mathcal{O}(R_{\mathrm{typ}}/R)}\label{eq:estimate_error1}.
\end{align}
As discussed above, the magnitude of the higher-order term $\mathcal{O}(R_{\mathrm{typ}}/R)$ cannot be determined from the observables because it is a model-dependent quantity. To estimate the uncertainty of $X$ with the higher-order terms, we define 
\begin{align}
 \xi \equiv R_{\mathrm{typ}}/R >0.
\end{align}
In the application to hadrons, $R_{\mathrm{typ}}$ is determined by the lowest mass of the meson mediating the interaction as $R_{\rm typ}=1/m_{\mathrm{typ}}$. In specific model calculations, $R_{\rm typ}$ can be determined, e.g., by the spatial extent of the potential, or by the inverse of the momentum cutoff. Using this quantity $\xi$, together with the observables $a_0$ and $E_h$, we determine the upper and lower boundaries of the compositeness as
\begin{align}
X_{\mathrm{u}} = \frac{a_0/R+ \xi}{2-a_0/R - \xi},\hspace{3ex}X_{\mathrm{l}} =  \frac{a_0/R- \xi}{2-a_0/R + \xi}.\label{eq:ul_limit_X}
\end{align}
Namely, the uncertainty band of the compositeness is estimated by $X_{\mathrm{l}}<X <X_{\mathrm{u}}$. We examine the validity of this error evaluation using solvable models in Sect.~\ref{sec:model}.

\section{Weak-binding relation for the quasibound state}\label{sec:quasibound}
\subsection{Effective field theory}
In this section, we extend the weak-binding relation to the quasibound state which has a decay mode.
In the first part of this section, we show the detailed derivation of the extended weak-binding relation written in Ref.~\cite{Kamiya:2015aea}.
We consider the coupled channel system with two scattering channels (channel~1 and channel~2) and a discrete channel.
There exists an unstable quasibound state near the threshold of channel~1, which can decay into channel~2.
As in Sect.~\ref{sec:bound} we assume that the interaction is of short range with typical length scale $R_{\mathrm{typ}}$.  

We consider the nonrelativistic EFT with the field $\psi_i$, $\phi_i$ and $B_0$ whose quantum numbers are the same as those of the hadrons of channels $i=1,2$ and the quasibound state, respectively. The Hamiltonian is given as
\begin{align}
H &= H_{\mathrm{free}} + H_{\mathrm{int}} =\int d^3\bm{x} (\mathcal{H}_{\mathrm{free}}+\mathcal{H}_{\mathrm{int}}), \\
	\mathcal{H}_{\mathrm
		{free}} &=\sum_{i=1,2}\frac{1}{2 M_i} \mathbf{\nabla} \psi_i^\dagger(\bm{x}) \cdot\mathbf{\nabla} \psi_i(\bm{x})  
	+\sum_{i=1,2}\frac{1}{2 m_i} \mathbf{\nabla} \phi_i^\dagger(\bm{x})  \cdot\mathbf{\nabla} \phi_i (\bm{x}) + \frac{1}{2M_{0}} \mathbf{\nabla}  B_0^\dagger(\bm{x})  \cdot{\mathbf \nabla} B_0(\bm{x}) \notag \\
	&\quad-\omega_\psi \psi_2^\dagger (\bm{x}) \psi_2(\bm{x}) -\omega_\phi \phi_2^\dagger(\bm{x})  \phi_2(\bm{x}) +\omega_0 B_0^\dagger (\bm{x}) B_0 (\bm{x}). \\
	\mathcal{H}_{\mathrm{int}} &=\sum_{i=1,2} g_{0,i} \left( B_0^\dagger (\bm{x}) \psi_i(\bm{x}) \phi_i (\bm{x}) + \phi_i^\dagger(\bm{x}) \psi_i^\dagger (\bm{x}) B_0 (\bm{x}) \right) 
	+ \sum_{i,j=1,2} v_{0,ij} \psi_j^\dagger (\bm{x})\phi_j^\dagger (\bm{x}) \phi_i(\bm{x}) \psi_i(\bm{x}),
\end{align}
where $g_{0,i}$ and $v_{0,ij}$ are the coupling constants of three- and four-point interactions, respectively.

Using Noether's theorem, the following particle numbers are conserved:
\begin{align}
	N_{\mathrm{I}}&\equiv N_{\psi_1} + N_{\psi_2} + N_{B_0} \label{eq:conservation-quasi1}\ ,\\
	N_{\mathrm{II}}&\equiv N_{\psi_1} - N_{\phi_1} \label{eq:conservation-quasi2}\ ,\\
	N_{\mathrm{III}}&\equiv N_{\psi_2} - N_{\phi_2} \label{eq:conservation-quasi3}.
\end{align}
From now on, we consider the sector of $N_{\mathrm{I}}= 1$, $N_{\mathrm{II}}=N_{\mathrm{III}}=0$, which contains the scattering states of channel $i=1,2$ and the discrete state of $B_0$.
Corresponding eigenstates of the free Hamiltonian are given by
\begin{align}
	|\bm{p}_i\rangle & \equiv \frac{1}{\sqrt{ V_p}}\tilde{\psi}_i^\dagger(\bm{p}) \tilde{\phi}_i^\dagger(-\bm{p})|0\rangle \hspace{3ex} (i = 1,2),\\
	|B_0\rangle &\equiv \frac{1}{\sqrt{V_p}}\tilde{B}_0^\dagger(\bm{0})|0\rangle,
\end{align}
where we define the vacuum $|0\rangle$ as in Eq.~\eqref{eq:vacuum}.
These states are again normalized and orthogonal to each other. The eigenenergies are $\bm{p}^2/(2\mu_i)-\omega\delta_{i,2}$ and $\omega_0$, where we define the reduced mass of channel $i$ as $\mu_i = M_i m_i/(M_i+m_i)$ and $\omega \equiv \omega_\psi +\omega_\phi$. The threshold energies of channel 1 and 2 are at $E=0$ and $E=-\omega$, respectively.  
From the particle number conservation laws, the completeness relation of this sector is
\begin{align}
\sum_{i=1,2}\int \frac{d^3\bm{p}}{(2\pi)^3}  |\bm{p}_i\rangle\langle\bm{p}_i| + |B_0\rangle\langle B_0| =1\label{eq:completeness2}.
\end{align}

\subsection{Scattering amplitude}

Next we derive the scattering amplitude of the $\psi_1\phi_1$ channel.
As in Sect.~\ref{sec:bound}, from the coupled-channel Schr\"odinger equation, we derive the effective single-channel equation for the wave function in channel 1.
The effective interaction which includes the contribution of scattering state 2 $|\bm{p}_2 \rangle$ and discrete state $|B_0\rangle$ is given by
\begin{gather}
v_1^{\mathrm{eff}}(E) \equiv  v_{11}(E) + \frac{v_{12}^2(E)}{[G_2(E)]^{-1} - v_{22}(E)}, \label{eq:v-eff-quasi}\\
v_{ij}(E) \equiv  v_{0,ij} + \frac{g_{0,i}g_{0,j} }{E-\omega_0},\quad 
G_i(E) \equiv \int \frac{d^3\bm{p}}{(2\pi)^3}    \frac{1}{E- p^2/(2\mu_i) + \delta_{i,2}\omega+i0^+} .
\end{gather}
The Lippmann--Schwinger equation for channel~1 becomes  
\begin{align}
t(E) = v_1^{\mathrm{eff}}(E) +v_1^{\mathrm{eff}}(E) G_1(E)t(E).
\end{align}
Solving this equation, we obtain the $T$ matrix $t(E)$ and the scattering amplitude $\mathcal{F}(E)$ of channel 1 as
\begin{align}
	t(E) = \frac{1}{[v_1^{\mathrm{eff}}(E)]^{-1}-G_1(E)} ,\quad
\mathcal{F}(E)= -\frac{\mu_1}{2\pi} \frac{1}{[v_1^{\mathrm{eff}}(E)]^{-1} -G_1(E)}.
\end{align}
The pole condition is given by 
\begin{align}
[ v_1^{\mathrm{eff}}(E_h)]^{-1} -G_1(E_h)=0 \label{eq:cond-quasi},
\end{align}
where $E_h$ denotes the eigenenergy.
From now on, we consider an unstable eigenstate $|\Psi_{\mathrm{QB}} \rangle$ with complex 
$E_h$, which follows the Schr\"odinger equation:
\begin{align}
H |\Psi_{\mathrm{QB}} \rangle &= E_h |\Psi_{\mathrm{QB}} \rangle,\\
	|\Psi_{\mathrm{QB}} \rangle &=  \sum_{i=1,2}\int \frac{d^3\bm{p}}{(2\pi)^3}\chi_i(\bm{p})|\bm{p}_i\rangle +c |B_0\rangle .
\end{align} 
The scattering length of channel~1, $a_0 = -\mathcal{F}(0)$, becomes
\begin{align}
	a_0 =&  \frac{\mu_1}{2\pi}\frac{1}{\left[v_1^{\mathrm{eff}}(0)\right]^{-1} - G_1(0) }. \label{eq:scat-amp-quasi}
\end{align}
The scattering length is in general complex, because $v_1^{\mathrm{eff}}(E)$ contains $G_2(E)$ which is complex at $E=0$. 

\subsection{Compositeness}
The unstable state cannot be normalized by the ordinary condition Eq.~\eqref{eq:normalization}.
To normalize the eigenstate $|\Psi\rangle$, here we introduce the Gamow state~\cite{Berggren:1968zz,Hokkyo1965} defined as
\begin{align}
	|\tilde{\Psi} \rangle \equiv  \sum_{i=1,2}\int \frac{d^3\bm{p}}{(2\pi)^3}\chi_i^*(\bm{p})|\bm{p}_i\rangle +c^* |B_0\rangle \label{eq:def-Gamow}.
\end{align}
Then the normalization condition is given by 
\begin{align}
	\langle \tilde{\Psi} | \Psi \rangle = \sum_{i=1,2}\int \frac{d^3\bm{p}}{(2\pi)^3}\chi_i^2(\bm{p}) + c^2 = 1\label{eq:Gamow-normalization-quasi}.
\end{align}
With the completeness relation Eq.~\eqref{eq:completeness2}, we define the compositeness $X_i$ ( the elementariness $Z$) as the sum of projections to the scattering state $|\bm{p}_i\rangle$ (the projection to the discrete state $|B_0\rangle$):
\begin{align}
	X_i \equiv \int \frac{d^3\bm{p}}{(2\pi)^3} \langle \tilde{\Psi} | \bm{p}_i\rangle  \langle \bm{p}_i | \Psi \rangle =  \int \frac{d^3\bm{p}}{(2\pi)^3} \chi_i^2(\bm{p}) ,\quad  
	Z \equiv \langle \tilde{\Psi}| B_0 \rangle \langle B_0 | \Psi \rangle = c^2 \label{eq:def-X-Z-2channel},
\end{align} 
where $X_i$ and $Z$ are complex because the weight $\chi_i(\bm{p})$ and $c$ are complex numbers in general.
With the normalization condition Eq.~\eqref{eq:Gamow-normalization-quasi}, we obtain the sum rule for $X_i$ and $Z$:
\begin{align}
	X_1 +X_2 +Z =1 \label{eq:X1+X2+Z}.
\end{align}
Unlike the bound state case, we cannot consider complex $X_i$ and $Z$ as the probabilities because they do not satisfy  condition~\eqref{eq:X-Z-in-0-to-1}.
We will discuss the interpretation of complex $X_i$ and $Z$ in Sect.~\ref{subsec:interpretation}. In the rest of this subsection, we derive an expression for the compositeness with the experimental observables.

From the Schr\"odinger equation for the quasibound state, we obtain 
\begin{gather}
	\chi_i(\bm{p})= \frac{1}{E_h - p^2/(2\mu_i) + \delta_{i,2}\omega}\sum_j v_{ij}(E_h)\alpha_j \label{eq:chi-multi},\\
c =\sum_{i=1,2}\frac{g_{0,i}}{E_h -\omega_0}\alpha_i,\quad
\alpha_i \equiv \int \frac{d^3\bm{p}}{(2\pi)^3}\chi_i(\bm{p}),
\end{gather}
which ensure that $X_i$ and $Z$ are written as 
\begin{gather}
	X_i =  -G_i^\prime(E_h) \beta_i^2,\quad	Z =  -\sum_{i,j}G_i(E_h)v_{ij}^{\prime}(E_h)G_j(E_h) \beta_i\beta_j,\quad 
		\beta_i \equiv  \sum_j v_{ij}(E_h) \alpha_j.\label{eq:X_i,Z}
\end{gather}
As we see in the next subsection, 
$X_1$ can be model-independently determined but the individual contributions of $X_2$ and $Z$ cannot be determined in the weak-binding limit.
Eliminating $\beta_1$ from Eq.~\eqref{eq:X_i,Z} with Eq.~\eqref{eq:X1+X2+Z}, we can write $X_1$ using the effective interaction and the derivative of the loop function of channel 1: 
\begin{align}
	X_1 = \frac{G_1^\prime(E_h)}{G_1^\prime(E_h)- \left[1/v_1^{\mathrm{eff}}(E_h)\right]^\prime }\label{eq:X-W-2channel}.
\end{align}
We note that this representation can be obtained with replacements $G\rightarrow G_1$ and $v\rightarrow v^{\mathrm{eff}}_1$ in  Eq.~\eqref{eq:X-Z-bound}.

\subsection{Weak-binding relation}\label{subsec:weak-limit-quasi}
Here we consider the case that the eigenenergy lies near the threshold energy of channel 1 and far from that of channel 2.
In this case, we can derive the weak-binding relation between the compositeness of channel 1 and the observables.

As in Eq.~\eqref{eq:def-R}, we define the quantity $R$ as
\begin{align}
R = \frac{1}{\sqrt{-2\mu_1 E_h}}.
\end{align}
In contrast to the bound state case, $R$ is now complex.
In addition, there is another length scale $l$ related to the difference between the threshold energies $\omega$ as
\begin{align}
l \equiv \frac{1}{\sqrt{2\mu_1 \omega}}.
\end{align}
Because $l$ is determined by the kinematics of the system, it is independent of the interaction range $R_{\mathrm{typ}}$.
In the weak-binding limit, the absolute value of $E_h$ is so small that $R$ is much larger than the interaction range $R_{\mathrm{typ}}$ and the length scale $l$: $|R_{\mathrm{typ}}/R| \ll 1$, $|l/R|\ll 1$. 

Expanding the scattering length in Eq.~\eqref{eq:scat-amp-quasi} by $1/|R|$ using 
Eqs.~\eqref{eq:X-W-2channel} and \eqref{eq:cond-quasi}, we obtain 
\begin{align}
a_0 =\frac{\mu_1}{2\pi} \left\{ \frac{E_h G_1^{\prime}(E_h)}{X_1}  +\Delta\left[v^{\mathrm{eff}}_1\right]^{-1} + \Delta G_1\right\}^{-1} \label{eq:a0-exp-2channel},
\end{align}
where $\Delta\left[v^{\mathrm{eff}}_1\right]^{-1}$ and $\Delta G_1$ are the higher-order terms of the expansion as in Eq.~\eqref{eq:Delta-v-G}.
The terms $G_1^\prime(E_h)$ and $\Delta G_1$ are estimated as in Eqs.~\eqref{eq:G-bibun-exp} and \eqref{eq:Delta-exp}, respectively.
Thus the term in Eq.~\eqref{eq:a0-exp-2channel} that we have to estimate is $\Delta\left[v_1^{\mathrm{eff}}\right]^{-1} $.

By assuming again that the eigenstate pole is in the convergence region of the effective range expansion,
the expansion of $\Delta\left[v_1^{\mathrm{eff}}\right]^{-1}$ starts from $E_h^2$.
Thus the leading terms of $\Delta\left[v_1^{\mathrm{eff}}\right]^{-1}$ must be
\begin{align}
	\Delta\left[v_1^{\mathrm{eff}}\right]^{-1} \propto \frac{1}{R^4}  \label{eq:Delta-v1eff}.
\end{align}
From Eq.~\eqref{eq:v-eff-quasi}, the explicit expression of $[v_{1}^{\mathrm{eff}}(E)]^{-1}$ is given as 
\begin{align}
	\left[v_{1}^{\mathrm{eff}}(E)\right]^{-1} = \frac{1-G_2(E)v_{22}(E)}{\left(1-G_2(E)v_{22}(E)\right)v_{11}(E)+G_2(E)v_{12}^2(E)},
\end{align}
which includes the contribution from the decay channel through $G_2$.
From this expression, we find that $\Delta\left[v_1^{\mathrm{eff}}\right]^{-1}$ includes higher derivatives of $v_{ij}(E)$ or $G_2(E)$.
The derivatives of $v_{ij}(E)$ are estimated as $(R_{\mathrm{typ}}/R)^n$.
The higher derivatives of $G_2(E)$ are written as 
\begin{align}
G_2(E) =&\frac{\mu_2}{\pi^2}\left[-\Lambda +\sqrt{2\mu_2 (E+\omega)} \mathrm{arctanh}\left(\frac{\Lambda}{\sqrt{2\mu_2 (E+\omega)}}\right)\right] \notag \\ 
 =&\frac{\mu_2}{\pi^2}\left[-\Lambda - \frac{i\pi}{2} \sqrt{2\mu_2 (E+\omega)} +\frac{2\mu_2 (E+\omega)}{\Lambda} + \frac{(2\mu_2(E+\omega))^2}{3\Lambda^3}+ \cdots\right], \\
E_h^n\frac{d^{n}G_2}{dE^n}(E_h)
=&  \frac{\mu_2 }{\pi^2} \Big[i\pi\frac{(-1)^{n-1}\left(2n-3\right)!!}{2^{n+2}}\sqrt{2\mu_2}\frac{E_h^n}{(E_h+\omega)^{(2n-1)/2}} + \frac{(2\mu_2 E_h)^n}{\Lambda^{2n+1}}(1 + \cdots)\Big]\notag \\
=& \frac{\mu_1}{R} \left[\mathcal{O}\left( \left(\frac{l}{|R|}\right)^{2n-1}\right) + \mathcal{O} \left(\left(\frac{R_{\mathrm{typ}}}{|R|}\right)^{2n-1}\right) \right],
\end{align}
with the assumption $\mu_2/\mu_1\sim \mathcal{O}(1)$.\footnote{It is a basic assumption in EFT that there is no accidental fine tuning that causes an order difference in the parameters.}
Thus we can obtain the following expression for $\Delta\left[v_1^{\mathrm{eff}}\right]^{-1}$: 
\begin{align}
\Delta\left[v_1^{\mathrm{eff}}\right]^{-1} &=\sum_{n=0}^3\frac{\mu_1}{R}\mathcal{O}\left(\frac{R_{\mathrm{typ}}^nl^{3-n}}{|R|^3}\right)
\end{align}
We note that, in contrast to the expansion of $G(E)$ in Eq.~\eqref{eq:Delta-exp}, $G_2(E)$ can be safely expanded due to the presence of the energy difference $\omega$.
By substituting this into Eq.~\eqref{eq:a0-exp-2channel}, the weak-binding relation for the quasibound state is derived as
\begin{align}
a_0 &= \frac{\mu_1}{2\pi} \Bigg\{ \frac{\mu_1}{4\pi X_1 R} + \frac{\mu_1}{4\pi R} + \frac{\mu_1}{R}{\mathcal O}\left(\left|\frac{R_{\mathrm{typ}}}{R}\right|\right) \notag\\
&\quad+ \frac{\mu_1}{R}\left[\mathcal{O}\left(\left|\frac{R_{\mathrm{typ}}}{R}\right|^{3}\right)+\mathcal{O}\left(\left|\frac{R_{\mathrm{typ}}}{R}\right|^{2}\right)\mathcal{O}\left(\left|\frac{l}{R}\right|\right) +\mathcal{O}\left(\left|\frac{R_{\mathrm{typ}}}{R}\right|\right)\mathcal{O}\left(\left|\frac{l}{R}\right|^2\right)+ \mathcal{O}\left(\left|\frac{l}{R}\right|^{3}\right) \right]\Biggr\}^{-1}  \notag \\
&=  R \left\{\frac{2X_1}{1+X_1} + {\mathcal O}\left(\left|\frac{R_{\mathrm{typ}}}{R}\right|\right) +  \mathcal{O}\left(\left|\frac{l}{R}\right|^{3}\right) \right\} \label{eq:comp-rel-quasi}.
\end{align}
The first and second term take the same form as that in the weak-binding relation for the bound state.
The contribution from the decay channel is expressed by the third term, which can be neglected when the eigenstate satisfies $|l/R|^{3} \ll 1$.
By neglecting this term, this relation reduces to the same form as Eq.~\eqref{eq:comp-rel-bound}.
When $|R|$ is large enough to neglect the second and third terms
in Eq. (77), the compositeness $X_1$ of the quasibound state
can be determined only by the observable quantities.
The conclusion is valid not only for the quasibound state with $\text{Re }E_h<0$ but also for that with $\text{Re }E_h>0$ because we have not assumed the sign of $\text{Re }E_h$ in the above derivation. 

Thus, as in the stable bound state case, the compositeness of the quasibound state with small $E_h$ can be model-independently determined by the experimental observables.
But unlike the bound state case, the compositeness is obtained as a complex number.

\subsection{Interpretation of complex compositeness}\label{subsec:interpretation}
Now we discuss the interpretation of the complex compositeness, utilizing some concrete examples. Because we can determine the quantity $X_1$  model-independently from the relation Eq.~\eqref{eq:comp-rel-quasi}, we write $X=X_1$ and $Z=1-X_1$ from here on.
 
For the bound state, the compositeness defined with Eq.~\eqref{eq:def-X} is a real number. With the two relations~\eqref{eq:X+Z} and \eqref{eq:X-Z-in-0-to-1}, we can interpret this quantity as the probability of finding the composite component.
On the other hand, for the unstable state, condition \eqref{eq:X-Z-in-0-to-1} is not satisfied. So we cannot interpret the complex $X$ as a probability.

To illustrate the problem, let us introduce three examples of compositeness.
Suppose that we obtain the compositeness of a quasibound state $X = 0.8-i0.2$ (case~I), $X=1.8-i0.1$ (case~II) and  $X=0.8-i0.9$ (case~III). The corresponding values of $Z$ are summarized in Table~\ref{tab:X-example}.
\begin{table}
\caption{Examples of the values of $X$ and $Z$, and the corresponding values of $\tilde{X}$, $\tilde{Z}$, $U$ with Eq.~\eqref{eq:kaisyaku2} }
\label{tab:X-example}
\begin{center}
\begin{tabular}{c|cc|ccc}
\hline
 Case   &$X$&$Z$&$\tilde{X}$ &$\tilde{Z}$&$U$\\ \hline 
     I    &$0.8-i0.2$&$\phantom{-}0.2+i0.2$&$0.8$&$0.2$&$0.1$ \\
     II   &$1.8-i0.1$&$-0.8+i0.1$&$1.0$&$0.0$&$0.8$ \\
     III  &$0.8-i0.9$&$\phantom{-}0.2+i0.9$&$0.6$&$0.4$&$0.6$ \\ \hline
\end{tabular}
\end{center}
\end{table}

In previous works, several interpretations of complex compositeness are proposed. 
For example, the absolute value of $X$ is considered as a probability in Ref.~\cite{Aceti:2012dd}.
With this method, however, $|X|$ becomes larger than unity when the cancelation in $X+Z$ is large, as in cases (II) ($|X|=1.8$) and (III) ($|X|=1.2$).\footnote{In Ref.~\cite{Guo:2015daa}, it is argued that $|X|<1$ holds if the Laurent series around the resonance pole converges in a region of the real axis. 
As we will show below, the difference between $|X|$ and $\tilde{X}$ we propose becomes small in the narrow width limit, and we have the condition $\tilde{X}<1$ by definition. 
This is in accordance with Ref.~\cite{Guo:2015daa} because the convergence of the Laurent series can be justified in the narrow width limit.}
In Ref.~\cite{Aceti:2014oma}, it is claimed that the real part of $X$ can be considered as an amount of the composite component based on the relation $\mathrm{Re}\hspace{1ex} X+ \mathrm{Re}\hspace{1ex} Z = 1$.
In this prescription, $\mathrm{Re} \int d^3\bm{p} \psi(\bm{p})^2$ for  case~(II) would be negative, so as studied in Ref.~\cite{Aceti:2014oma}, this  cannot be taken as a probability.

We now analyze the examples in more detail. 
For case~(I), the real part satisfies $0< \mathrm{Re} \hspace{1ex} X<1$ and the imaginary part is small. 
For cases~(II) and (III), there is a large cancelation in the sum of the real part and imaginary part to satisfy the sum rule of $X+Z=1$.
This difference may be attributed to the magnitude of the decay width of the quasibound state.
When the quasibound state has a small decay width, its wave function is expected to be similar to that of a bound state~\cite{Sekihara:2014kya}.
In this case, the value of the compositeness should be similar to the bound state case. The probability of finding the scattering state in the quasibound state in case~(I) is then expected to be $\sim80\%$. 
Hence the structure of the narrow width state is reflected in $X$ and $Z$. 
On the other hand, for cases~(II) and (III), the corresponding bound state wave function does not exist. Therefore we cannot discuss the internal structure of the eigenstate from complex $X$, in contrast to case (I).

\begin{figure}[htbp]
  \begin{center}
   \includegraphics[width=90mm]{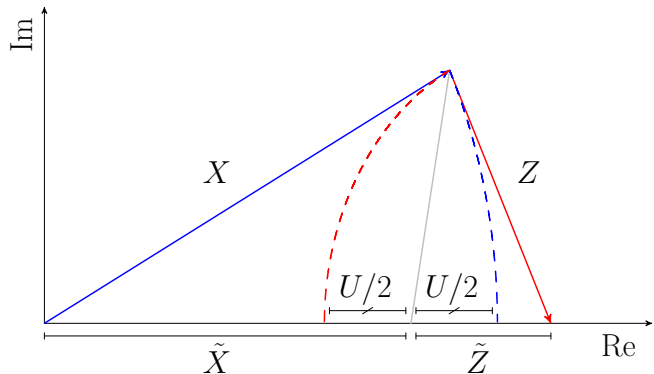}
  \end{center}
  \caption{
    Geometric illustration of $\tilde{X}$, $\tilde{Z}$ and $U$ defined in Eq.~\eqref{eq:kaisyaku2}.
    }
    \label{fig:Xtilde}
\end{figure}

In Ref.~\cite{Kamiya:2015aea}, we introduced the new real quantities $\tilde{X}$, $\tilde{Z}$ and $U$ based on this observation.
To interpret $\tilde{X}$ and $\tilde{Z}$ as the probabilities of finding the scattering state and the other state, the following conditions should be satisfied:
\begin{itemize}
\setlength{\leftskip}{2cm}
 \item[Condition (1)] $\tilde{X},\tilde{Z} \in [0,1]$;  
 \item[Condition (2)] $\tilde{X} + \tilde{Z} = 1$. 
\end{itemize}
As we can see from the above examples, a reasonable interpretation is not always guaranteed.
When the amount of the cancelation in $X+Z$ is large, 
the probabilistic interpretation is not appropriate.
Only when the cancelation is small can we use the probabilistic interpretation. 
To measure the uncertainty of the interpretation, we introduce the quantity $U$ which satisfies the following conditions: 
\begin{itemize}
\setlength{\leftskip}{2cm}
 \item[Condition~(3)]  When there is no cancelation in $X+Z$, then $U=0$, $\tilde{X}=X$ and  $\tilde{Z}=Z$;
 \item[Condition~(4)] $U$ increases as the cancelation in $X+Z$ becomes large.
\end{itemize}
We define these three quantities as
\begin{align}
\tilde{X} \equiv \frac{1 - |Z| + |X|}{2},\quad \tilde{Z} \equiv \frac{1 - |X| + |Z|}{2},\quad 
    U \equiv |Z| +|X| -1 \label{eq:kaisyaku2} ,
\end{align}
which can be calculated from $X$ and $Z$.\footnote{In Ref.~\cite{Berggren:1970}, the probability of an uncertain identification of the state $c_n$ is defined with complex overlap of the wave function $p_n$. We define $U$ motivated by this prescription.}
From the triangle inequalities $|X|+|Z| \geq 1$, $|X|+1 \geq |Z|$ and $|Z|+1 \geq |X|$, we can verify that the quantities $\tilde{X}$, $\tilde{Z}$, and $U$ satisfy the four conditions.
A geometric illustration of the definition of these quantities is given in Fig.~\ref{fig:Xtilde}. 
The relation $X+Z=1$ can be expressed in the complex plane as $(\re X,\im X)+(\re Z,\im Z)=(1,0)$. 
In this figure, we can regard $U$ as the difference between $|X|$ and $1-|Z|$ on the real axis. The quantity $\tilde{X}=1-\tilde{Z}$ is defined by taking the middle point of $(|X|,0)$ and $(1-|Z|,0)$. 
From this observation, it is reasonable to consider $\pm U/2$ as the uncertainty of the probability $\tilde{X}$. We emphasize that this uncertainty comes from the complex nature of the expectation values of the unstable particle. We thus call $U/2$ the uncertainty of the interpretation.
We show these quantities for the examples of $X$ in Table~\ref{tab:X-example}.
For case~(I), $U$ is small enough to regard the value of 
$\tilde{X}=0.8$ as a probability, which implies the structure is dominated by the composite state. 
For the other cases, $U/2$ is larger than $1/2$ and the large uncertainty prevents us from employing the probabilistic interpretation.
In this way, we can quantitatively discuss the structure of the unstable states by this interpretation with $\tilde{X}$ and $U$.

We note that Eq.~\eqref{eq:kaisyaku2} is not the only definition to satisfy the above four conditions.
For example, another definition is proposed in Ref.~\cite{Sekihara:2015gvw}.\footnote{A nice feature of the definition in Ref.~\cite{Sekihara:2015gvw} is that it is generalizable to a system with multiple scattering channels. We note, however, that the only quantity we can model-independently determine is the compositeness of the nearest channel to the pole. Thus the present formulation is sufficient  to interpret the model-independently determined compositeness.}
As shown in Ref.~\cite{Sekihara:2015gvw}, the difference between the definitions should be small when $U$ is small.
When $U$ is small the other interpretations of the complex compositeness, $|X|$ and $\text{Re }X$, also give a similar result to $\tilde{X}$.
In fact, the differences between these expressions reduce to
\begin{align}
\left||X| -\tilde{X}\right| &=\frac{U}{2}, \\
\left|\text{Re } X -\tilde{X}\right|&= \Bigl||X|-|Z|\Bigr| \frac{U}{2} \leq \frac{U}{2},
\end{align}
where we have used $-1\leq|Z| -|X|\leq 1$.\footnote{With this inequality, we can also show that the difference between $\tilde{X}$ and that in Ref.~\cite{Sekihara:2015gvw} is smaller than $U/2$.}
Though $|X|$ and $\text{Re }X$ do not always satisfy condition (1), the difference from $\tilde{X}$ is of the order of $U$. 
Thus in the case of small $U$, they also give a reasonable estimation of the compositeness, which is consistent with $\tilde{X}$.

\subsection{Error evaluation of compositeness}\label{sec:error}

Finally we construct a method to evaluate the error of the compositeness of the quasibound states. In contrast to the stable bound states in Sect.~\ref{sec:bound}, the compositeness $X$ and the higher-order terms are all complex numbers for the quasibound states. Because the probabilistic interpretation is applied to the real-valued $\tilde{X}$, here we consider its upper and lower boundaries $\tilde{X}_{\mathrm{u}}$ and $\tilde{X}_{\mathrm{l}}$.

To estimate the effect of the higher-order terms, we first introduce a complex quantity $\xi_{c}$ in the expression of the compositeness as
\begin{align}
X = \frac{a_0/R + \xi_{c}}{2-a_0/R-\xi_{c}}\label{eq:estimate_error_quasi}.
\end{align}
In the present case, $\xi_{c}$ is made of two components $\mathcal{O}(|R_{\mathrm{typ}}/R|)$ and $\mathcal{O}(|l/R|^3)$. Both terms are  in general complex with an unknown relative phase. As a conservative error estimation, we allow $\xi_{c}$ to vary in the region
\begin{align}
   |\xi_{c}| & \leq |R_{\mathrm{typ}}/R|+|l/R|^3 
   \label{eq:condition} .
\end{align}
In other words, the largest magnitude of $\xi_{c}$ is determined when two terms are coherently added. We then evaluate $\tilde{X}$ with Eq.~\eqref{eq:kaisyaku2} by varying $\xi_{c}$ with Eq.~\eqref{eq:condition} being the constraint. Denoting the maximum (minimum) value of $\tilde{X}$ as $\tilde{X}_{\mathrm{u}}$ ($\tilde{X}_{\mathrm{l}}$), we consider the uncertainty band of $\tilde{X}$ as $\tilde{X}_{\mathrm{u}}<\tilde{X} < \tilde{X}_{\mathrm{l}}$.\footnote{We note that $\tilde{X}_{\mathrm{u}/\mathrm{l}}$ is not always given by $\xi_{c}$ with the inequality~\eqref{eq:condition} being saturated. For instance, if $|a_{0}/R|<|R_{\mathrm{typ}}/R|+|l/R|^3$, $\tilde{X}_{\mathrm{l}}=0$ is given by $\xi_{c}=-a_{0}/R$ whose magnitude is smaller than $|R_{\mathrm{typ}}/R|+|l/R|^3$.}

\section{Weak-binding relation with CDD pole contribution}\label{sec:CDD}

In the derivation of the weak-binding relation in the previous sections, we have assumed the convergence of the effective range expansion (ERE) at the eigenenergy $R_{\mathrm{eff}} \lesssim R_{\mathrm{typ}}$ above Eq.~\eqref{eq:estimation-v}.
While the ERE is a general expression of the near-threshold amplitude, its convergence region does not always reach the eigenenergy.
 In this section, we extend the weak-binding relation to the case where the ERE does not work well. For simplicity, here we consider the stable bound state. The generalization to the unstable quasibound state is straightforward as in Sect.~\ref{sec:quasibound}.

The validity of the ERE is related to the magnitude of $R_{\mathrm{eff}}$ in Eq.~\eqref{eq:eff-exp}.
When $R_{\mathrm{eff}}$ is large, the convergence of the expansion is limited to the small energy region.
The most drastic case occurs when the CDD (Castillejo--Dalitz--Dyson) pole lies near the threshold energy.
Because the CDD pole is defined as the pole of the inverse scattering amplitude $\mathcal{F}(E)^{-1}$,
the ERE converges only in the region $|E|<|E_c|$, where $E_c$ is the closest CDD pole to the threshold.
When the CDD pole is close to the eigenenergy, $|E_h|\sim|E_c|$,
the description by the ERE is not appropriate at the eigenenergy, and then we cannot use the weak-binding relation~\cite{Weinberg:1965zz}.
The effect of the CDD pole is discussed in relation to near-threshold states and the compositeness in Refs.~\cite{Baru:2010ww,Guo:2016wpy}.

\subsection{Weak-binding expansions of compositeness }\label{subsec:model-dependency} 

We first show an alternative derivation of the weak-binding relation~\eqref{eq:comp-rel-bound}, by paying attention to the cutoff dependence of the components. In the new derivation, the expansion of $R_{\mathrm{typ}}/R$ and $R_{\mathrm{eff}}/R$ can be performed separately, so that the extension of the weak-binding relation is achieved by improving the expansion of $R_{\mathrm{eff}}/R$.
We define the coupling constant $g$ between the scattering state and the bound state from the residue of the $T$ matrix:
\begin{align}
g^2 \equiv & \lim_{E\rightarrow E_h}(E-E_h)t(E)\label{eq:g-def}.
\end{align}
With Eq.\eqref{eq:T-mat}, this becomes
\begin{align}
g^2=\left. \frac{E -E_h}{v^{-1}(E) - G(E)}\right|_{E=E_h} = -\frac{1}{-\left[1/v(E_h)\right]^\prime +G^\prime(E_h)}. \label{eq:g-vG}
\end{align}
Substituting this expression to Eq.~\eqref{eq:X-Z-bound}, the compositeness can be written with $g$ and $G^\prime$ as  
\begin{align}
X =& -g^2 G^\prime(E_h) \label{eq:X-gG} .
\end{align}
This expression has been used in many works~\cite{Hyodo:2011qc,Aceti:2012dd,Sekihara:2014kya,Guo:2015daa}.

Now we examine the cutoff-dependence of the coupling constant $g$.
To do this, we slightly change the cutoff as $\Lambda \rightarrow \Lambda +\delta \Lambda$. 
Because the scattering amplitude is the observable, we have to reproduce the original $t(E)$ modifying the effective interaction. Then the modified effective interaction $v_{\delta\Lambda}(E)$ should satisfy
\begin{align}
t(E) = \frac{1}{v^{-1}(E) - G(E;\Lambda)} = \frac{1}{v^{-1}_{\delta\Lambda}(E) - G(E;\Lambda+\delta\Lambda)},
\end{align}
where we write the cutoff of the loop function explicitly.
Solving this equation for $v_{\delta\Lambda}(E)$, we obtain
\begin{align}
v_{\delta\Lambda}(E) &= \left[ v^{-1}(E) + G(E;\Lambda+\delta\Lambda) -G(E;\Lambda)\right]^{-1}.
\end{align}
The scattering amplitude with $\Lambda +\delta \Lambda$ gives the new bound state condition $v^{-1}_{\delta\Lambda}(E_h)G(E_h;\Lambda+\delta\Lambda)=1$ with the same eigenenergy $E_h$. Then we can see that $g^2$ is invariant under the change of the cutoff:
\begin{align}
g^2 &\rightarrow -\left[ v_{\delta\Lambda}^{-2}(E_h) v_{\delta\Lambda}^\prime(E_h) +G^\prime(E_h;\Lambda+\delta\Lambda)\right]^{-1} \notag\\
      & = -\left[ v_{\delta\Lambda}^{-2}(E_h) v^2_{\delta\Lambda}\left(v^{-2}(E_h)v^\prime(E_h) + G(E_h;\Lambda+\delta\Lambda) - G(E_h;\Lambda)\right) +G^\prime(E_h;\Lambda+\delta\Lambda)\right]^{-1} \notag\\
      & = -\left[ v^{-2}(E_h)v^\prime(E_h) +G^\prime(E_h;\Lambda)\right]^{-1}.
\end{align}
In Eq.~\eqref{eq:X-gG}, the cutoff dependence of the compositeness comes only from the derivative of the loop function, because $g$ has no cutoff dependence.
In other words, the expansion in powers of $R_{\mathrm{typ}}/R$ is only applicable for $G^\prime(E_h)$ in Eq.~\eqref{eq:X-gG}, because the length scale $R_{\mathrm{typ}}$ is related to the cutoff as $R_{\mathrm{typ}}\sim 1/\Lambda$.
From Eq.\eqref{eq:G-bibun-exp}, the derivative of the loop function is given as
\begin{eqnarray}
G^\prime(E_h) =-\frac{\mu^2 R}{2\pi} \left[1+\mathcal{O}\left(\frac{R_{\mathrm{typ}}}{R}\right)\right] \label{eq:G-bibun-exp-rewritten}.
\end{eqnarray}
From here on, we derive the weak-binding relation~\eqref{eq:comp-rel-bound} by applying the ERE to the coupling constant $g$.
Substituting the general form of the scattering amplitude  Eq.~\eqref{eq:scat-delta}, 
$g^2$ can be expressed as
\begin{align}
g^2 = \lim_{E\rightarrow E_h}(E-E_h)t(E)
       = \left.\frac{2\pi}{\mu^2}\left[\frac{1}{p}\frac{d(p\cot\delta)}{dp} -i\frac{1}{p}\right]^{-1}\right|_{p=i/R}.
\end{align}
Using the effective range expansion, this coupling constant is written using $R$ and $r_{\mathrm{e}}$:
\begin{align}
g^2 &= \frac{2\pi}{\mu^2}\frac{1}{ R-r_{\mathrm{e}} +R\mathcal{O}((R_{\mathrm{eff}}/R)^3)} \label{eq:g-exp}.
\end{align}
With this expression and Eq.~\eqref{eq:G-bibun-exp-rewritten}\, the compositeness in Eq.~\eqref{eq:X-gG} reduces to
\begin{align}
X    &= \frac{1}{1-\frac{r_{\mathrm{e}}}{R} +\mathcal{O}\left(\left(\tfrac{R_{\mathrm{eff}}}{R}\right)^3\right)}\left[1 + \mathcal{O}\left(\frac{R_{\mathrm{typ}}}{R}\right)\right] \label{eq:comp-rel-re-R} .
\end{align}
As in the previous section, assuming $R_{\mathrm{eff}} \lesssim R_{\mathrm{typ}}$, we can derive the weak-binding relation
\begin{align}
X &= \frac{R}{R-r_{\mathrm{e}} }\left[1+\mathcal{O}\left(\frac{R_{\mathrm{typ}}}{R}\right)\right]\label{eq:X-R-re-2},
\end{align}
which is expressed using $R$ and $r_{\mathrm{e}}$.
Eliminating $r_{\mathrm{e}}$ using the bound state condition in the effective range expansion;
\begin{align}
-\frac{1}{a_0} - \frac{r_{\mathrm{e}}}{2R^2} +\frac{1}{R}+\frac{1}{R}\mathcal{O}\left(\left(\frac{R_{\mathrm{eff}}}{R}\right)^3\right)=0,\label{eq:bound-cond-ERE}
\end{align}
Eq.~\eqref{eq:X-R-re-2} reduces to the previous relation Eq.~\eqref{eq:comp-rel-bound}.

In this derivation, we notice that there are two expansions, both of which contain the higher-order correction terms.
Equation~\eqref{eq:comp-rel-re-R} is derived with the following assumptions:
\begin{itemize}
\item[(i)] $R$ is sufficiently larger than the typical range scale of the interaction: $R_{\mathrm{typ}}/R \ll 1$,
\item[(ii)] the convergence region of ERE reaches the bound state pole: $(R_{\mathrm{eff}}/R)^3 \ll 1$.
\end{itemize}
Assumption (i) is related to the expansion of the derivative of the loop function and the higher-order term of $\mathcal{O}(R_{\mathrm{typ}}/R)$ depends on the cutoff $\Lambda$.
With assumption (ii), the higher-order term of $\mathcal{O}((R_{\mathrm{eff}}/R)^3)$ arises from the expansion of $g^{-2}$. This higher-order term is not related to the cutoff because the coupling constant $g$ is cutoff independent.
If we assume 
\begin{itemize}
\item[(iii)] $R_{\mathrm{eff}} \lesssim R_{\mathrm{typ}}$,
\end{itemize}
assumption (ii) follows from assumption (i) and we obtain the weak-binding relation Eq.~\eqref{eq:X-R-re-2} which is equivalent to Eq.~\eqref{eq:comp-rel-bound}.
Without this assumption (iii), the two expansions are independent of each other and these higher-order terms should be considered separately.

\subsection{Improvement by effective range expansion}\label{mod-ERE}

Equation \eqref{eq:comp-rel-re-R} is obtained by approximating $p\cot\delta$ up to the $p^2$ term in the ERE.
To approximate $p\cot\delta$, here we take the fourth-order term of the expansion: 
\begin{align}
p\cot\delta = -\frac{1}{a_0} + \frac{r_{\mathrm{e}}}{2}p^2 + \frac{v}{4}p^4+\mathcal{O}(R_{\mathrm{eff}}^5p^6)\label{eq:mod-e.r.e}.
\end{align}
Substituting Eq.~\eqref{eq:mod-e.r.e} into Eq.~\eqref{eq:g-exp}, we obtain the expression for $g^2$ with $R$, $r_{\mathrm{e}}$ and $v$.
Then the modified weak-binding relation is obtained as
\begin{align}
X  =\left[1-\frac{r_{\mathrm{e}}}{R}+\frac{v}{R^3} +\mathcal{O}\left(\left(\frac{R_{\mathrm{eff}}}{R}\right)^5\right)+\mathcal{O}\left(\frac{R_{\mathrm{typ}}}{R}\right)\right]^{-1}\label{eq:comp-rel-mod-ERE1}. 
\end{align}
When $v/R^3$ is of order $\mathcal{O}((R_{\mathrm{eff}}/R)^3)$, 
this expression reduces to Eq.~\eqref{eq:comp-rel-re-R}. 
Including the term of $v/R^3$, the estimation of the compositeness is improved. 
This relation can be rewritten using the condition of the bound state:
\begin{align}
-\frac{1}{a_0}-\frac{r_{\mathrm{e}}}{2}\frac{1}{R^2}+ \frac{v}{4}\frac{1}{R^4} + \frac{1}{R}+\frac{1}{R}\mathcal{O}((R_{\mathrm{eff}}/R)^5)=0. \label{eq:bound-cond-mod-e.r.e}
\end{align}
By eliminating $v$ with Eq.~\eqref{eq:bound-cond-mod-e.r.e},
the compositeness is given by $R$, $a_0$, and $r_{\mathrm{e}}$ as 
\begin{align}
X  =\left[\frac{4R}{a_0}+\frac{r_{\mathrm{e}}}{R}-3 +\mathcal{O}\left(\left(\frac{R_{\mathrm{eff}}}{R}\right)^5\right)+\mathcal{O}\left(\frac{R_{\mathrm{typ}}}{R}\right)\right]^{-1}\label{eq:comp-rel-mod-ERE2}.
\end{align}
When $R$ satisfies $(R_{\mathrm{eff}}/R)^5\ll 1 $ and $R_{\mathrm{typ}}/R\ll 1$, we can neglect the higher-order terms and calculate the compositeness from $a_0$, $r_{\mathrm{e}}$ and, $E_h$. 
In Eq.~\eqref{eq:comp-rel-mod-ERE2}, the contribution from the higher-order terms of the effective range expansion is included in $R$ through condition~\eqref{eq:bound-cond-mod-e.r.e}. 

\subsection{Improvement by Pad\'e approximant}\label{subsec:Pade}

When a CDD pole lies near the threshold, the convergence region of the effective range expansion may not reach the bound state pole.
Here we use the Pad\'e approximant method to describe $p\cot\delta$
\begin{align}
p\cot\delta = \frac{b_0 + b_1 p^2}{1 + c_1p^2} + \mathcal{O}\left( R_{\mathrm{Pad\acute{e}}}^5p^6\right)\label{eq:Pade},
\end{align}
where $R_{\mathrm{Pad\acute{e}}}$ is the length scale characterizing this expansion.
With this method, we can describe the scattering amplitude which has a CDD pole at $p = \pm i/\sqrt{c_1}$. This is because $f(p)\rightarrow 0$ when $p\cot\delta \rightarrow \infty$. 
The threshold parameters are related to the expansion coefficients as:
\begin{align}
a_0 = -\frac{1}{b_0} ,\quad
r_{\mathrm{e}} = 2(b_1-b_0c_1)\label{eq:a0-re-Pade}.
\end{align}
Substituting Eq.~\eqref{eq:Pade} into Eq.~\eqref{eq:g-exp}, we obtain an expression for the coupling constant:
\begin{align}
g^2 =& - \frac{2\pi p}{\mu^2}\left.\left\{\frac{2b_1p (1+c_1p^2 ) - 2c_1p(b_0 + b_1 p^2)   }{\left(1+c_1p^2 \right)^2} -i+\mathcal{O}\left(R_{\mathrm{Pad\acute{e}}}^5p^5 \right)\right\}^{-1}\right|_{p=i/R}.
\end{align}
By using this equation, the compositeness is given by
\begin{align}
X=\left[\frac{2(b_1-c_1b_0)R^2}{(c_1-R^2)}-1 + \mathcal{O}\left(\left(\frac{R_{\mathrm{Pad\acute{e}}}}{R}\right)^5\right)+\mathcal{O}\left(\frac{R_{\mathrm{typ}}}{R}\right)\right]^{-1}\label{eq:comp-rel-Pade1}.
\end{align}
When $c_1 \rightarrow 0$, this expression reduces to Eq.~\eqref{eq:comp-rel-re-R} with $b_1 = r_{\mathrm{e}}/2$. Owing to the nonzero value of $c_1$, the contribution of the CDD pole is included in the estimation of the compositeness.  
Although the compositeness is expressed with three coefficients and $R$ in this expression, 
the coefficients $b_0$ and $b_1$ appear in the combination of $b_1-b_0c_1 =r_{\mathrm{e}}/2$. 
Thus these three independent quantities are rewritten with $a_0$, $r_{\mathrm{e}}$, and $R$ using Eq.~\eqref{eq:a0-re-Pade} and the condition of the bound state:
\begin{align}
\left(b_0 - b_1 \frac{1}{R^2} \right) +\frac{1}{R}\left(1 - c_1\frac{1}{R^2} \right) \left[1 + \mathcal{O}\left(\left(\frac{R_{\mathrm{Pad\acute{e}}}}{R}\right)^5\right)\right] =0 \label{eq:pole-cond-Pade}
\end{align}
Thus the extended weak-binding relation becomes
\begin{align}
X=\left[1 - \frac{4R(a_0-R)^2}{a_0^2r_{\mathrm{e}}}+ \mathcal{O}\left(\left(\frac{R_{\mathrm{Pad\acute{e}}}}{R}\right)^5\right)+\mathcal{O}\left(\frac{R_{\mathrm{typ}}}{R}\right)\right]^{-1}\label{eq:comp-rel-Pade2}.
\end{align}
By neglecting the two higher-order terms, the compositeness can be determined with the three observables $a_0$, $r_{\mathrm{e}}$, and $E_h$.
In the derivation of this equation, we have employed the Pad\'e approximant, which enables us to include the CDD pole contribution.
Thus the compositeness of the near-threshold bound state with a nearby CDD pole can be determined using this extended relation.

\section{Model calculations}\label{sec:model}

From the discussion in Sect.~\ref{sec:bound}, we can estimate the compositeness of the weak-binding state with the relation~\eqref{eq:comp-rel-bound}.
In Sect.~\ref{sec:CDD}, we introduced two more weak-binding relations by improving the estimation of the coupling constant.
By using models where the scattering amplitude and the exact value of the compositeness of the bound state can be numerically calculated, we study the validity of the determination of the compositeness by the weak-binding relations qualitatively.
Furthermore, we investigate the validity of the error evaluation with Eq.~\eqref{eq:ul_limit_X}.
For simplicity, we consider the stable bound state case.

\subsection{Square-well potential model}

We consider $s$-wave scattering by the square-well potential, 
which is described by the following radial Schr\"odinger equation:
\begin{gather} 
\left[ -\frac{1}{2\mu}\frac{d^2}{dr^2} + V(r) -E\right] u(r) = 0 ,\quad
V(r) = \begin{cases}
	-V_0 & (r < b )\\
	0    & (r \ge b)
\end{cases} ,\label{eq:Schrodinger-eq-potential}
\end{gather}
where $\mu=469.5 \physdim{MeV}$ is the reduced mass, and $V_0$ and $b$ are the depth and the range of the potential. We consider the case where one bound state appears in this scattering problem. For a given $b$, the lower limit of the depth to have a single bound state is $V_0^c = \pi^2/(8\mu b^2)$.
The eigenstates of the free Hamiltonian of the corresponding quantum field theory consist only of the scattering states. 
Thus the compositeness of the bound state definitely becomes unity.
We first examine the estimation of the compositeness by the weak-binding relation:
\begin{align}
X_{a_0,R} = \frac{a_0}{2R-a_0},\label{eq:comp-rel-bound-rewritten}
\end{align}
which is obtained by neglecting the higher-order term of $\mathcal{O}(R_\mathrm{typ}/R)$.
Here we continuously increase the potential depth $V_0$ from the lower limit $V_0^c$ keeping the value of the range $b=1 \physdim{fm}$.
Then the eigenenergy $E_h$ increases from $0$ with $V_0$.
In this model, the typical range scale of the interaction is $R_{\mathrm{typ}}  \sim  b =1\physdim{fm}$.
Given the potential parameters, the scattering length is given by $a_0 = b-(2\mu V_0)^{-1/2}\tan[(2\mu V_0)^{1/2}b]$ and we can numerically determine the eigenenergy.
Then the compositeness can be estimated by Eq.~\eqref{eq:comp-rel-bound-rewritten}.

\begin{figure}[t]
	\begin{center}
		\includegraphics[width = 10 cm ]{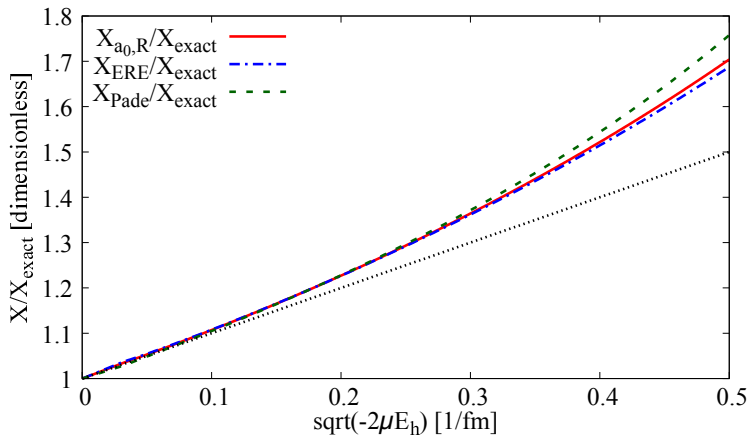}
		\caption{
		The estimated compositeness normalized by the exact value in the square-well potential model. The results with Eqs.~\eqref{eq:comp-rel-bound-rewritten},  \eqref{eq:comp-rel-mod-ERE-rewritten} and  \eqref{eq:comp-rel-Pade-rewritten} are denoted by the solid line, the dash-dotted line, and the dashed line, respectively. The dotted line denotes $b/R+1$}.
		\label{fig:X-fix-b}
	\end{center}
\end{figure}
We show the ratio of the estimation to the exact value $X_{\mathrm{exact}}=1$, as a function of $\sqrt{-2\mu E_h}$ by the solid line in Fig.~\ref{fig:X-fix-b}.
Because of the neglected higher-order term, the estimated value is larger than $1$.
For comparison, we plot $b/R + 1$ with a dotted line.
We can see that the amount of the error of the estimation is of the order of $b/R$, as expected from $\mathcal{O}(R_{\mathrm{typ}}/R)$.
We discuss the effect of the higher-order terms in Sect.~\ref{subsec:model_error} in more detail.
An accurate estimation is possible in the weak-binding limit $E_h\rightarrow0$.
To see the validity of the effective range expansion at the pole position, we calculate the dimensionless quantity
\begin{eqnarray}
\delta \equiv \left|-\frac{R}{a_0 } - \frac{r_{\mathrm{e}}}{2R}  +1\right|\label{eq:delta},
\end{eqnarray}
which indicates the size of the higher-order terms $\mathcal{O}((R_{\mathrm{eff}}/R)^3)$ in Eq.~\eqref{eq:bound-cond-ERE}.
In the plotted eigenenergy region, we obtain $\delta < 0.02$, which implies that the ERE converges well and
the value of the higher-order terms $\mathcal{O}((R_\mathrm{eff}/R)^3)$ related to assumption (ii) discussed in Sect.~\ref{subsec:model-dependency} is small.
Therefore the error of the estimation comes mainly from the  $\mathcal{O}(R_\mathrm{typ}/R)$ term related to assumption (i).
This can be confirmed by estimating the compositeness with the extended relations:
\begin{align}
X_{\mathrm{ERE}} &=\left[\frac{4R}{a_0}+\frac{r_{\mathrm{e}}}{R}-3\right]^{-1}\label{eq:comp-rel-mod-ERE-rewritten}, \\
X_{\mathrm{Pad\acute{e}}}&=\left[1 - \frac{4R(a_0-R)^2}{a_0^2r_{\mathrm{e}}}\right]^{-1}\label{eq:comp-rel-Pade-rewritten}.
\end{align}
We show the results of the estimation normalized by the exact value in Fig.~\ref{fig:X-fix-b}.
We see that there is no large difference among the three estimations. This is because the ERE converges well in the plotted eigenenergy region, and the deviation from unity is dominated by $\mathcal{O}(R_{\mathrm{typ}}/R)$.

\subsection{Contact interaction model}
In this subsection, we consider a case with nonzero elementariness using the field theoretical model with contact interaction as introduced in Sect.~\ref{sec:bound}.
We consider the Hamiltonian~\eqref{eq:Hamiltonian-bound} with the sharp cutoff $\Lambda$ as the model of the scattering problem, not as an effective description of some underlying theory.
Then we discuss the CDD pole contribution in the estimation of the compositeness. 
Given the parameters in the Hamiltonian ($v_0$, $g_0$, $\omega_0$) and the cutoff $\Lambda$, we can calculate the scattering length and the effective range from the scattering amplitude~\eqref{eq:amp} at $E=0$.
The eigenenergy is related to these parameters by the pole condition~\eqref{eq:pole-condition}.
The exact value of the compositeness in this model $X_{\mathrm{exact}}$ is given by Eq.~\eqref{eq:X-Z-bound}. 

\begin{figure}[t]
  \begin{center}
   \includegraphics[width=140mm]{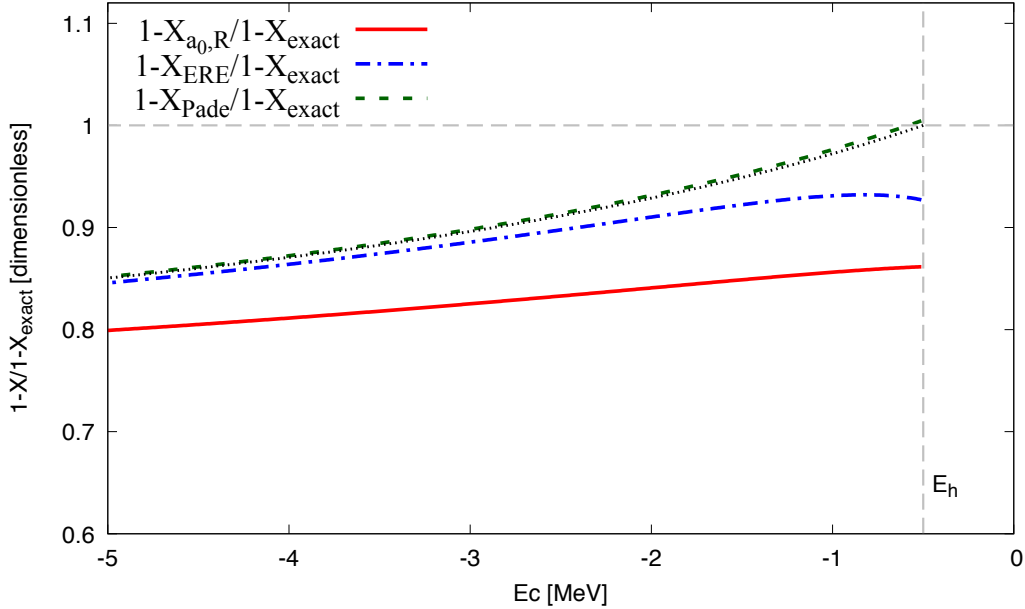}
  \end{center}
 \caption{The estimated elementariness $1-X$ normalized by the exact value with $\omega_0 = 5\physdim{MeV}$ and $E_h=-0.5 \physdim{MeV}$.
  The notation for the compositeness is the same as Fig.~\ref{fig:X-fix-b}.
  The results with Eqs.~\eqref{eq:comp-rel-bound-rewritten},  \eqref{eq:comp-rel-mod-ERE-rewritten}, and  \eqref{eq:comp-rel-Pade-rewritten} are denoted by the solid line, dash-dotted line, and dashed line, respectively.
  The dotted line represents $(1-g^2\mu^2R/(2\pi))/(1-X_{\mathrm{exact}})$. 
 }
  \label{fig:case1_0.5}
\end{figure}

From Eqs.~\eqref{eq:v-eff-bound} and \eqref{eq:amp}, the amplitude has a single CDD pole at $E_c = \omega_0 -g_0^2/v_0$.
Adjusting the parameters in the Hamiltonian, we can let the CDD pole lie near the eigenenergy, ($E_c\sim E_h$). 
We set the reduced mass as $\mu = 469.5 \physdim{MeV}$, the cutoff as $\Lambda = 140\physdim{MeV}$, and vary $E_h$ and $E_c$ by changing the values of $g_0$, $v_0$, and $\omega_0$.
The coupling constant $g_0$ must be real so that the Hamiltonian is Hermitian.
Because $g_0^2$ is written using Eqs.~\eqref{eq:v-eff-bound} and \eqref{eq:pole-condition} as 
\begin{align}
g_0^2=\frac{(\omega_0 - E_c)(\omega_0-E_h)}{E_c-E_h}G^{-1}(E_h), \label{eq:g2}
\end{align}
and $G(E_h) < 0$ for $E_h<0$, then
$E_h$, $E_c$, and $\omega_0$ must satisfy one of the following three conditions:
\begin{itemize}
\setlength{\leftskip}{2cm}
\item[Case 1 :] $E_c < E_h<\omega_0$;
\item[Case 2 :] $\omega_0 <E_c < E_h$;
\item[Case 3 :] $E_h < \omega_0 < E_c$.
\end{itemize}
We will consider cases (1) and (2) to study the validity of the weak-binding relations when $E_c \sim E_h$.
From now on, we fix the values of $\omega_0$ and $E_h$, and continuously change the position of the CDD pole by the parameters.
We note that, in the limit of $E_c\rightarrow E_h$,
$g^2$ vanishes because the divergence of $dv^{-1}/dE|_{E=E_h}$ in Eq.~\eqref{eq:g-vG} is proportional to $1/\epsilon$ in the limit of $\epsilon\rightarrow 0$, where $\epsilon =E_c -E_h$.
This means that $X_{\mathrm{exact}}$ also vanishes  from Eq.~\eqref{eq:X-gG}. 
Then the estimated compositeness normalized with $X_{\mathrm{exact}}$ may diverge in this limit. 
To normalize the estimation with a nonzero quantity, here we use  the elementariness $1-X =Z$ normalized by $1-X_{\mathrm{exact}} = Z_{\mathrm{exact}}$ to quantify the validity of the estimations.

\begin{figure}[t]
  \begin{center}
   \includegraphics[width=140mm]{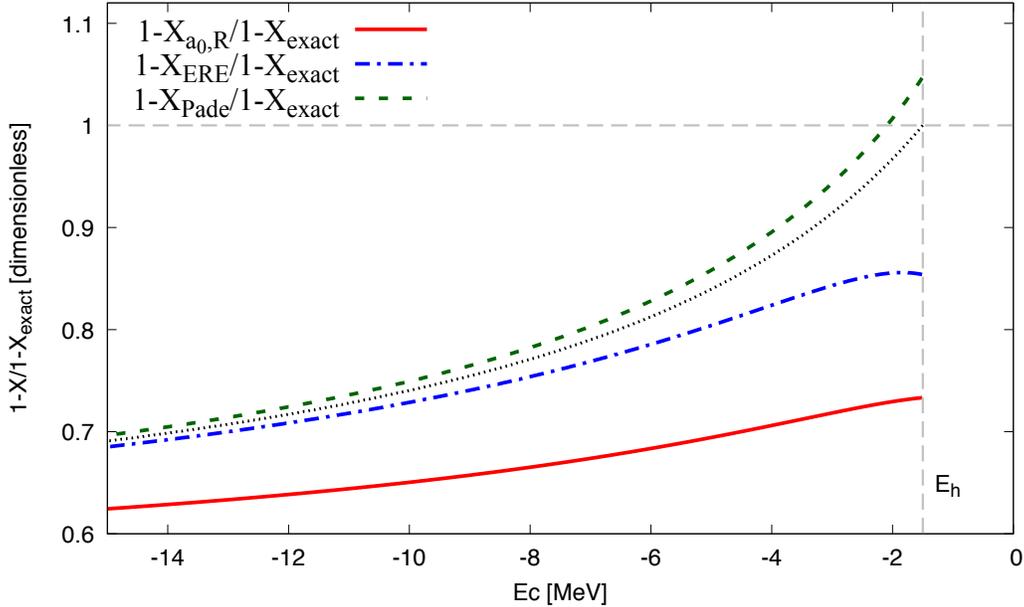}
  \end{center}
 \caption{The ratio of estimated elementariness $1-X$ to the exact value with $\omega_0=5\physdim{MeV}$ and $E_h=-1.5 \physdim{MeV}$.
 The notation for the elementariness is the same as Fig.~\ref{fig:case1_0.5}.
 }
  \label{fig:case1_1.5}
\end{figure}

\begin{figure}[htbp]
  \begin{center}
   \includegraphics[width=140mm]{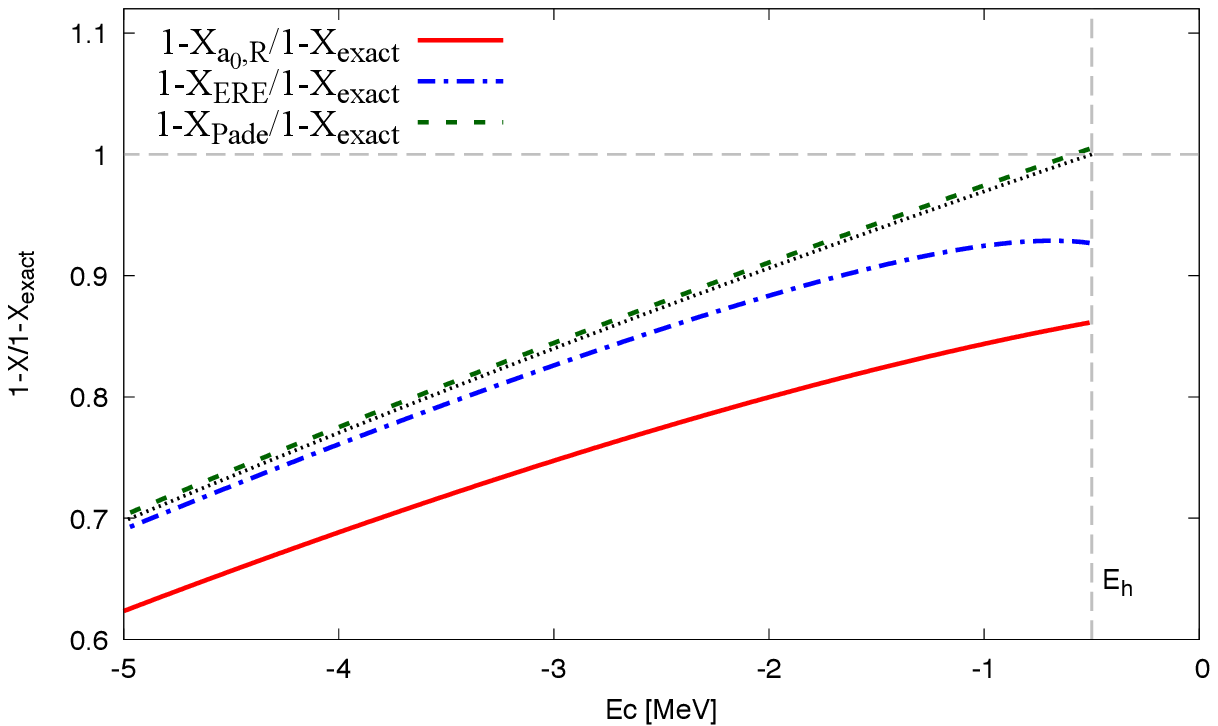}
  \end{center}
  \caption{
    The ratio of the estimated elementariness $1-X$ to the exact value for $E_h > \omega_0$ with $E_h=-0.5\physdim{MeV}$. The notation for the elementariness is the same as Fig.~\ref{fig:case1_0.5}
    }
    \label{fig:case2_0.5}
  \end{figure}

The estimated values of the normalized elementariness as functions of $E_c$ are shown in Fig.~\ref{fig:case1_0.5}  with $\omega_0 = 5\physdim{MeV}$ and $E_h = 0.5\physdim{MeV}$.
This corresponds to the small binding energy case because $R_{\mathrm{typ}}/R=0.15$, where $R_{\mathrm{typ}} \sim 1/\Lambda$.
The estimations with $X_{a_0,R}$ in Eq.~\eqref{eq:comp-rel-bound-rewritten}, $X_{\mathrm{ERE}}$ in Eq.~\eqref{eq:comp-rel-mod-ERE-rewritten} and $X_{\mathrm{Pad\acute{e}}}$ in Eq.~\eqref{eq:comp-rel-Pade-rewritten} normalized by the exact value are denoted by the solid line, the dash-dotted line and the dashed line, respectively. 
We see that the estimation by $X_{\mathrm{ERE}}$ is closer to the exact value than that by $X_{a_0,R}$.
This improvement is considered to originate in the inclusion of the higher-order terms of the effective range expansion.
The estimation by $X_{\mathrm{Pad\acute{e}}}$ reproduces the exact value in the limit $E_c\rightarrow E_h$ where the CDD pole lies in the proximity of the bound state.
Thus using the extended relation~\eqref{eq:comp-rel-Pade2}, we can estimate the compositeness with the CDD pole contribution.

One may wonder why the estimation with $X_{a_0,R}$ still reproduces the exact results with a 20 \% error in the region of $E_c\sim E_h$.
To understand this, we consider the compositeness that is obtained by only approximating $G^\prime(E_h)$ in Eq.~\eqref{eq:X-gG} as
\begin{align}
X \sim g^2\frac{\mu^2R}{2\pi}, \label{eq:X_g,exact}
\end{align}
with the exact value of the coupling constant $g$ being used.
This estimated value contains the higher-order terms of $R_{\mathrm{eff}}/R$, but not those of $R_{\mathrm{typ}}/R$.
We plot $(1-g^2\mu^2R/(2\pi))/(1-X_{\mathrm{exact}})$ in Fig.~\ref{fig:case1_0.5} with a dotted line. 
From this estimation, we find that the higher-order terms of $R_{\mathrm{typ}}/R$ are suppressed at $E_c\rightarrow E_h$.
This originates in the vanishing of $g^2$ in the limit of $E_c \rightarrow E_h$.
Considering this point, the estimation with Eq.~\eqref{eq:comp-rel-bound-rewritten} contains substantial error from the higher-order terms of $R_{\mathrm{eff}}/R$, and we should use the improved  Eq.~\eqref{eq:comp-rel-Pade-rewritten} at $E_c \sim E_h$.

The same calculation as in Fig.~\ref{fig:case1_1.5} is performed with $\omega_0=5\physdim{MeV}$ and $E_h =-1.5 \physdim{MeV}$ ($R_{\mathrm{typ}}/R = 0.26$).
The deviations from the exact value are larger than the previous ones.
This is because the higher-derivative term related to assumption (i), $\mathcal{O}(R_{\mathrm{typ}}/R)$, is large. 
We can confirm this fact from the estimation with Eq.~\eqref{eq:X_g,exact} shown by the dotted line.
The estimation by $X_{\mathrm{Pad\acute{e}}}$ works well in the region $E_c\sim E_h$ similarly to the previous calculation.
Finally we show the result of case 2, where $\omega_0=-45\physdim{MeV}$ and $E_h=-0.5\physdim{MeV}$, in Fig.~\ref{fig:case2_0.5}. 
We see that the qualitative conclusion is the same as the calculation of case (1) with $E_h = -0.5 \physdim{MeV}$.
The quantitative difference between the estimation from the calculation in Fig.~\ref{fig:case1_0.5} results again from the larger $\mathcal{O}(R_{\mathrm{typ}}/R)$ as indicated by the dotted line.

\subsection{Validity of the error evaluation}\label{subsec:model_error}

In Sect.~\ref{eq:errorbound}, we introduced a method to evaluate the uncertainty of the compositeness that comes from the higher-order correction terms in the weak-binding relation. The upper and lower boundaries of the uncertainties of $X$ are given in Eq.~\eqref{eq:ul_limit_X}. Here we check the validity of this method by comparing the compositeness, including uncertainties in the weak-binding relation, with the exact values in solvable models. We employ one square-well potential model and two contact interaction models.

\begin{figure}[htbp]
  \begin{center}
   \includegraphics[width=140mm]{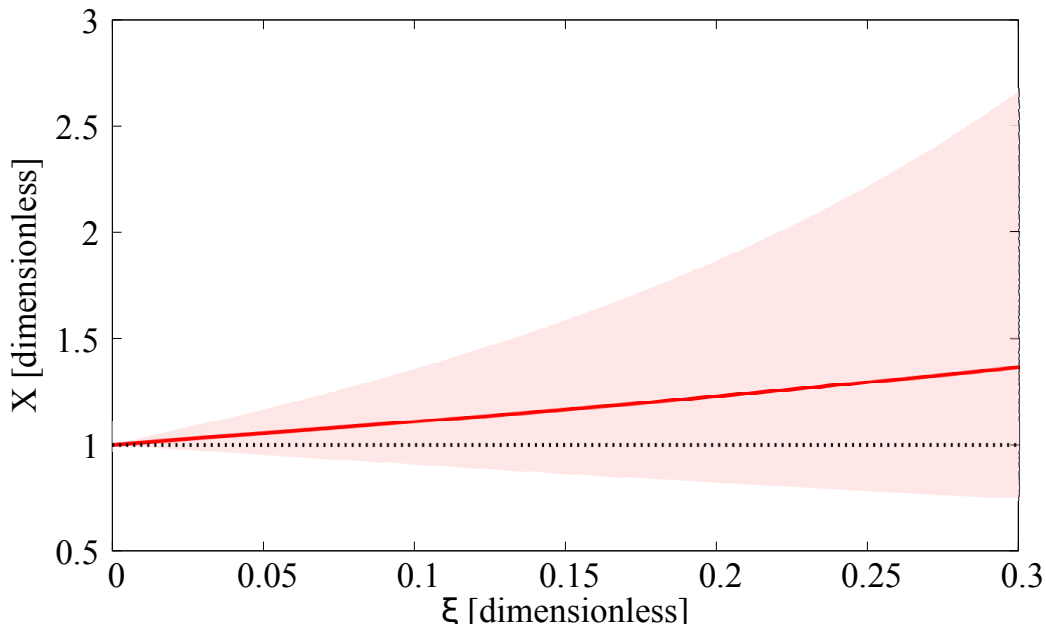}
  \end{center}
  \caption{
    Compositeness $X$ evaluated by the weak-binding relation~\eqref{eq:comp-rel-bound-rewritten} (solid line) with the uncertainty band given in Eq.~\eqref{eq:estimate_error1} (shaded area) as functions of $\xi = R_{\mathrm{typ}}/R=b\sqrt{-2\mu E_{h}}$ in the square-well potential model.
    The dotted line denotes the exact value of the compositeness $X_{\mathrm{exact}}=1$.}
    \label{fig:error_well}
  \end{figure}

We first consider the square-well potential model used in the previous subsection. As mentioned above, we vary the potential depth $V_{0}$ to change the eigenenergy $E_h$ with the fixed potential range $b=1$ fm. The reduced mass is set as $\mu=469.5 \physdim{MeV}$. In Fig.~\ref{fig:error_well}, we plot the central value of the compositeness (solid line) together with the uncertainty band (shaded area) in the weak-binding relation, as a function of $\xi = R_{\mathrm{typ}}/R=b\sqrt{-2\mu E_{h}}$. We observe in Fig.~\ref{fig:error_well} that, although the central value is larger than unity, the uncertainty band in the weak-binding relation covers the region $X<1$ that can be interpreted as probability. In particular, the exact value $X_{\rm exact}=1$ is included within the estimated uncertainty. We also see that the magnitude of the uncertainty reduces with $\xi$.
In the present case, we can qualitatively conclude that the bound state is dominated by the composite component, as long as $\xi\lesssim 0.3$.

\begin{figure}[htbp]
  \begin{center}
   \includegraphics[width=140mm]{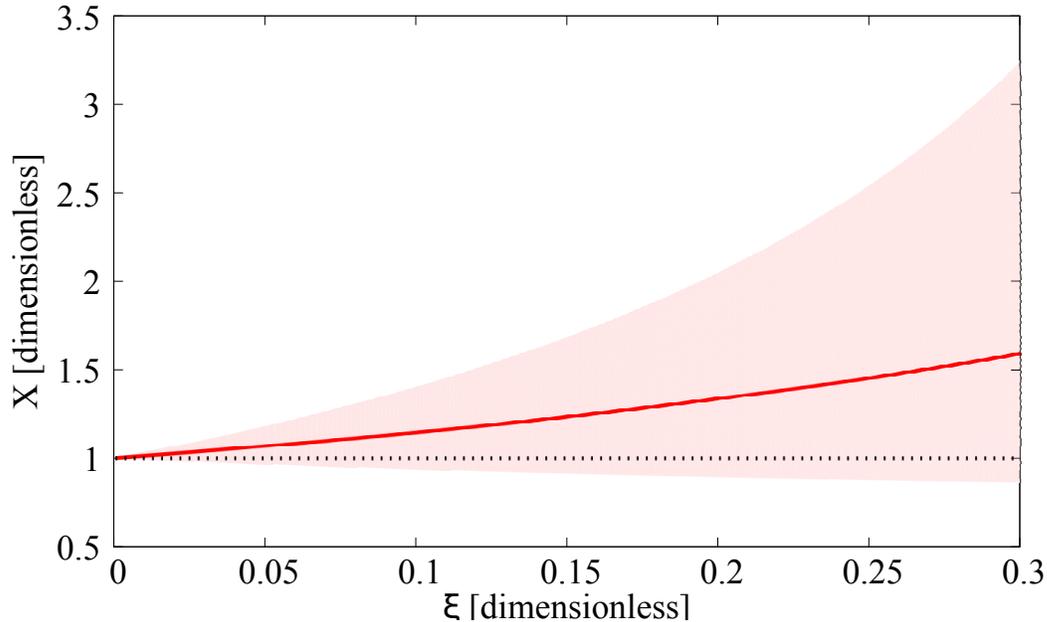}
  \end{center}
  \caption{Compositeness $X$ evaluated by the weak-binding relation~\eqref{eq:comp-rel-bound-rewritten} (solid line) with the uncertainty band given in Eq.~\eqref{eq:estimate_error1} (shaded area) as functions of $\xi = R_{\mathrm{typ}}/R=\sqrt{-2\mu E_{h}}/\Lambda$ in the contact interaction model without coupling to the bare field, $g_0=0$. The dotted line denotes the exact value of the compositeness $X_{\mathrm{exact}}=1$.}
    \label{fig:error_composite}
  \end{figure}
  
\begin{figure}[htbp]
    \begin{center}
     \includegraphics[width=140mm]{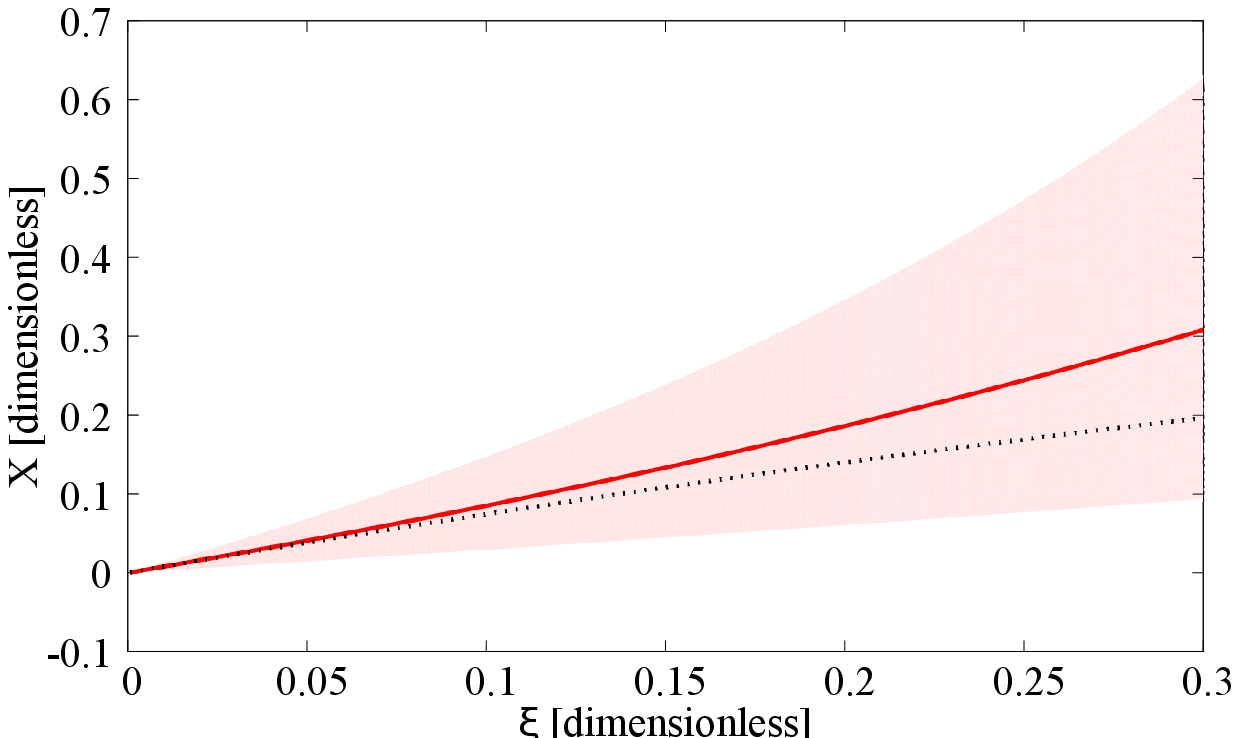}
    \end{center}
    \caption{Compositeness $X$ evaluated by the weak-binding relation~\eqref{eq:comp-rel-bound-rewritten} (solid line) with the uncertainty band given in Eq.~\eqref{eq:estimate_error1} (shaded area) as functions of $\xi = R_{\mathrm{typ}}/R=\sqrt{-2\mu E_{h}}/\Lambda$ in the contact interaction model without the four-point interaction, $v_0=0$ and $\omega_{0}=0$. The dotted line denotes the exact value of the compositeness $X_{\mathrm{exact}}$ evaluated by Eq.~\eqref{eq:X-Z-bound}.}
      \label{fig:error_elementary}
\end{figure}
 
Next, we consider a contact interaction model with Hamiltonian~\eqref{eq:Hamiltonian-bound} which possesses a compositeness-dominant bound state. For this purpose, we set $g_0=0$ so that the bare state $B_{0}$ decouples from the scattering state. By varying the strength of the four-point interaction $v_0$, we tune the eigenenergy $E_{h}$. Here we set the reduced mass $\mu = 469.5 \physdim{MeV}$ and the cutoff $\Lambda = 140\physdim{MeV}$. In Fig.~\ref{fig:error_composite}, we show the compositeness $X$ and its uncertainty band as functions of $\xi = R_{\mathrm{typ}}/R=\sqrt{-2\mu E_{h}}/\Lambda$. Because of the absence of the bare state, the exact value of the compositeness is given by $X_{\rm exact}=1$ as indicated by the dotted line. As in the case of the square-well potential, the error evaluation works well and the exact value is included in the uncertainty band.

Finally, we examine the case with an elementary-dominant bound state. To achieve this, we consider the model without the four-point interaction ($v_0=0$). As shown in Ref.~\cite{Hyodo:2014bda}, however, any $s$-wave bound state that couples with the scattering state becomes composite dominant in the weak-binding limit $|E_{h}|\to 0$. The only exception is the ``decoupling'' case where $g_{0}\to 0$ is taken simultaneously with $|E_{h}|\to 0$. An efficient way to realize this situation in the present model is to set $\omega_0 =0$. By varying the strength of the three-point interaction $g_{0}$ to change the eigenenergy $E_{h}$, the decoupling limit is automatically guaranteed. Again, we use $\mu = 469.5 \physdim{MeV}$ and $\Lambda = 140\physdim{MeV}$. The numerical results are shown in Fig.~\ref{fig:error_elementary}. Exact value of the compositeness $X_{\rm exact}$ is calculated by Eq.~\eqref{eq:X-Z-bound}. We see that $X_{\rm exact}$ vanishes at $\xi=0$ because of the decoupling limit. As expected, $X_{\rm exact}$ is well within the estimated uncertainty in this case also.

\subsection{Conclusion of the model calculations}
In this section, we have shown the results of the numerical model calculations to examine the weak-binding relations from various viewpoints. In all cases studied, we confirm that the compositeness can be appropriately determined in the weak-binding limit. With the help of the extended relation, the determination works well even if the CDD pole lies near the threshold.

On the other hand, it is also important to consider the case away from the strict weak-binding limit, because the effect of the higher-order correction terms is not negligible for actual hadron resonances. Based on the model calculations with both composite-dominant and elementary-dominant bound states, we find that the exact values are consistent with the uncertainty estimation developed in Sect.~\ref{sec:error}. Qualitative conclusions may be drawn if $\xi=R_{\rm typ}/R\lesssim 0.3$ in the cases studied.

At this point, it is instructive to compare the determination of the compositeness by the weak-binding relation with the evaluation of the compositeness at the pole position using Eq.~\eqref{eq:X-Z-bound}~\cite{Hyodo:2011qc,Aceti:2012dd,Sekihara:2014kya,Guo:2015daa}. In the latter method, we need a theoretical model to perform the analytic continuation of the scattering amplitude. A nice feature in this method is the absence of any uncertainty in the evaluation process, because the compositeness is evaluated exactly at the pole position. In addition, the method can be applied to the states in any partial waves with arbitrary eigenenergy. On the other hand, the result suffers from the model (or renormalization) dependence, as a consequence of the inherent off-shell nature of the compositeness. The magnitude of the model dependence is not under control within this approach. In practice, we need a sufficient amount of experimental data of the scattering to determine a reliable amplitude, which is in general difficult, e.g., in the heavy hadron sector.

In the weak-binding relation approach, the compositeness can be estimated with only a few observable quantities. Moreover, the result is model independent, as long as the weak-binding limit is concerned. Nevertheless, in practical applications, the estimation always contains uncertainty from the higher-order correction terms, because the compositeness is evaluated by the expansion of the amplitude. At the same time, we have to keep in mind that the applicability of the weak-binding relation is limited to $s$-wave near-threshold states. 

One may consider that the uncertainty in the weak-binding relation is a reflection of the model dependence of the compositeness, because the origin of the uncertainty is the model-dependent higher-order terms. As we have demonstrated here, it is possible to estimate the magnitude of this uncertainty/model-dependence for the weak-binding $s$-wave states, by using the weak-binding relation. We emphasize again that the model-independence of the compositeness of the near-threshold states is guaranteed only when the weak-binding relation is established. 

In this way, the two approaches are in some sense complementary. Motivated by these observations, in the next section, we discuss the determination of the compositeness of physical hadron resonances with the weak-binding relations, in comparison with the evaluations at the pole position in theoretical amplitudes.

\section{Applications to hadrons }\label{sec:application}
\subsection{Structure of $\Lambda(1405)$}
$\Lambda(1405)$ is the negative parity excited baryon with spin $1/2$ and isospin $I=0$~\cite{Agashe:2014kda}.
$\Lambda(1405)$ couples to the $\bar{K}N$ channel in $s$-wave and eventually decays into the $\pi\Sigma$ channel. 
Because the state lies near the $\bar{K}N$ threshold energy, 
we can study the $\bar{K}N$ compositeness of $\Lambda(1405)$ with the generalized weak-binding relation for quasibound states.
To evaluate the compositeness using the weak-binding relation, we need the $I=0$ scattering length of the $\bar{K}N$ channel and the eigenenergy of $\Lambda(1405)$. 
These quantities can be obtained by detailed fitting analysis of the experimental data in the $\bar{K}N$ threshold energy region. 
The most systematic analysis in the previous studies is performed by chiral SU(3) dynamics~\cite{Ikeda:2011pi,Ikeda:2012au,Mai:2012dt,Guo:2012vv,Mai:2014xna}. In these studies, $\Lambda(1405)$ is described by two resonance poles of the scattering amplitude in the complex energy plane.
We consider the $\bar{K}N$ compositeness of the state represented by the pole at higher energy because this can be regarded as the weakly bound state.\footnote{We do not consider the compositeness of the state associated with the lower-energy pole, because the weak-binding relation is derived for the closest pole to the threshold.}
In Table~\ref{tab:Lambda}, we show the sets of the scattering length $a_0$ and the eigenenergy of the higher pole state $E_h$, based on Refs.~\cite{Ikeda:2011pi,Ikeda:2012au,Mai:2012dt,Guo:2012vv,Mai:2014xna}.\footnote{We thank Jose Antonio Oller and Maxim Mai for correspondences.}
Because of the isospin symmetry breaking, the threshold energies and the reduced masses of the $\bar{K}^0n$ channel and the $K^- p$ channel are slightly different. 
We define the scattering length for the isospin $I=0$ channel as $a_0 = (f_{0,K^-p}(E=0)+f_{0,\bar{K}^0n}(E=0))/2$, where $f_{0,K^-p}$ and $f_{0,\bar{K}^0n}$ are the scattering amplitudes of $K^-p\rightarrow K^-p$ and $\bar{K}^0n\rightarrow \bar{K}^0n$, respectively, and the threshold energy $E=0$ is specified below for each set.
The scattering length of set 1 is calculated from the NLO amplitude of Refs.~\cite{Ikeda:2011pi,Ikeda:2012au} by using the isospin-averaged hadron masses at the isospin-averaged $\bar{K}N$ threshold energy.
Therefore we use the isospin-averaged mass of $\bar{K}$ and $N$ to determine the threshold energy and the reduced mass.
Set 3 is based on Fit II of Ref.~\cite{Guo:2012vv} with the same isospin-averaging procedure.
In the other analyses, the scattering length is calculated at the $K^{-}p$ threshold energy, so we use the threshold energy and reduced mass of the $K^-p$ channel.
Sets 2, 4, and 5 are based on Ref.~\cite{Mai:2012dt}, solution \#2 of Ref.~\cite{Mai:2014xna}, and solution \#4 of Ref.~\cite{Mai:2014xna}, respectively.
In Table~\ref{tab:Lambda}, the scattering length $a_0$ and the eigenenergy $E_h$ do not converge quantitatively even though the available data is reproduced at the level of $\chi^2/\text{d.o.f}\sim 1$ in all the analyses. 
We therefore employ the results of all the analyses to estimate the systematic error.

\begin{table}[bt]
		\caption{Properties and results for the higher-energy pole of $\Lambda (1405)$ quoted from Ref.~\cite{Kamiya:2015aea}: shown are the eigenenergy $E_{h}$, the $\bar{K}N(I=0)$ scattering length $a_{0}$, the $\bar{K}N$ compositeness $X_{\bar{K}N}$ and $\tilde{X}_{\bar{K}N}$ and the uncertainty of the interpretation $U$.}
		\label{tab:Lambda}
	\begin{center}
		\begin{tabular}{llllcl}
			\hline
              & $E_{h}\physdim{[MeV]}$ & $a_0 \physdim{[fm]} $ & $X_{\bar{K}N}$ & $\tilde{X}_{\bar{K}N}$ & $U/2$   \\  \hline
			 Set 1 \cite{Ikeda:2012au}  & $-10-i26$ & $1.39 - i 0.85$ 
			 & $1.2+i0.1$ & $1.0$ & $0.3$  \\ 
			 Set 2 \cite{Mai:2012dt}  & $-\phantom{0}4-i\phantom{0}8$ & $1.81-i0.92$ 
			 & $0.6+i0.1$ & $0.6$ & $0.0$ \\ 
			 Set 3 \cite{Guo:2012vv}  & $-13-i20$ & $1.30-i0.85$ 
			 & $0.9-i0.2$ & $0.9$ & $0.1$  \\
			 Set 4 \cite{Mai:2014xna}  & $\phantom{-0}2-i10$ & $1.21-i1.47$ 
			 & $0.6+i0.0$ & $0.6$ & $0.0$ \\ 
			 Set 5 \cite{Mai:2014xna}  & $-\phantom{0}3-i12$ & $1.52-i1.85$ 
			 & $1.0+i0.5$ & $0.8$ & $0.3$ \\ 
			 \hline
		\end{tabular} 
	\end{center}
\end{table}

We first estimate the magnitude of the higher-order terms in the weak-binding relation.
Using the eigenenergies in Table~\ref{tab:Lambda}, we find that the value of $R$ satisfies $|R| \gtrsim 1.5 \hspace{1ex}\mathrm{fm}$. 
The typical range scale of the hadron interaction can be estimated from the meson exchange mechanism.
The longest range hadronic interaction is mediated by the lightest meson $\pi$, which cannot be exchanged between $\bar{K}$ and $N$ because the three-point vertex of the pseudoscalar mesons is prohibited by parity conservation.
We therefore estimate the typical range scale of the $\bar{K}N$ interaction from the $\rho$ meson exchange interaction to obtain 
$R_{\mathrm{typ}} =1/m_\rho\sim 0.25 
\hspace{1ex}\mathrm{fm}$.\footnote{We do not use the $\sigma$ exchange to estimate the interaction range because the $\sigma$ meson has the broad width~\cite{Agashe:2014kda}.}
To estimate the length scale $l=1/\sqrt{2\mu \omega}$, we use the difference between 
the threshold energy of the $\bar{K}N$ channel and that of the $\pi\Sigma$ channel.
Taking $\omega = 104 \hspace{1ex}\mathrm{MeV}$, we obtain
$l=0.76\hspace{1ex}\mathrm{fm}$. 
Then we can see that the two higher-order terms in the weak-binding relation are small, $|R_{\mathrm{typ}}/R| \lesssim 0.17$ and $|l/R|^3 \lesssim 0.14$ for each set.

For each set, we have calculated the compositeness $X_{\bar{K}N}$, $\tilde{X}_{\bar{K}N}$ and the uncertainty of the interpretation $U$ with the weak-binding relation in Ref.~\cite{Kamiya:2015aea} by neglecting the higher-order terms.
The results are summarized in Table~\ref{tab:Lambda}.
In all sets, $U$ is not large and we can consider that
the uncertainty of the probability interpretation of $\tilde{X}_{\bar{K}N}$ is small. The evaluated value of $\tilde{X}_{\bar{K}N}$ is close to unity in all sets.
Thus we find that the structure of $\Lambda(1405)$ is dominated by the $\bar{K}N$ composite component.
The deviation of the values of the compositeness among the different sets comes from the difference between $a_0$ and $E_h$.
The uncertainty $U$ is determined for a given set of $a_0$ and $E_h$, which originates in the complex nature of the compositeness.
The former deviation can be reduced with the improvement of the analysis of the experimental data, and it will eventually vanish when the exact values of $a_0$ and $E_h$ are obtained.
However, the latter uncertainty is not necessarily small even for the exact values. In the present analysis, we conclude that the $\bar{K}N$ component of $\Lambda(1405)$ reaches $60\% \sim 100 \% $ of the total wave function, and the quantitative uncertainty will be squeezed in future improvement of the threshold parameters. Given the results in Table~\ref{tab:Lambda}, the associated value of $U$ is expected to be small.

\begin{figure}
 \begin{center}
 \includegraphics[width=60mm]{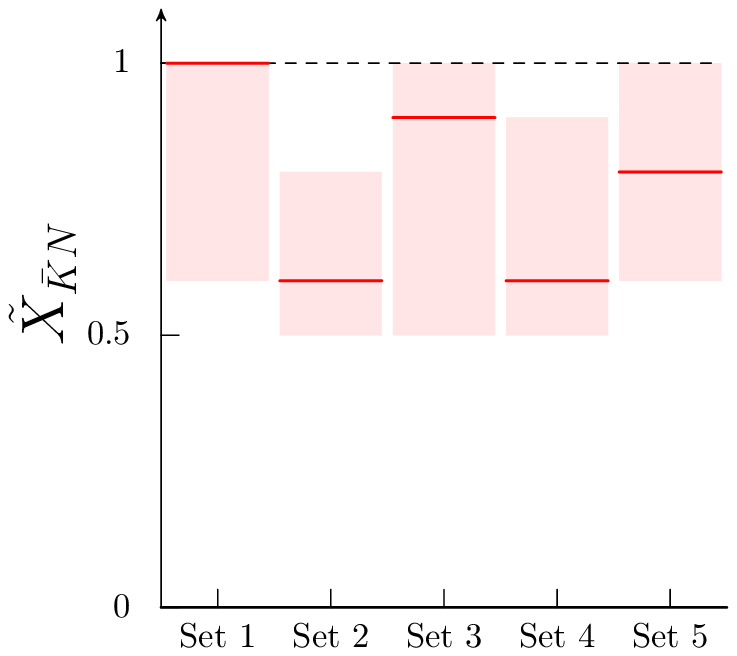}
   \caption{The results of error evaluation of the compositeness $\tilde{X}_{\bar{K}N}$ of  $\Lambda(1405)$.  The lines denote the central values and the shaded areas indicate the uncertainty bands.}
   \label{fig:error_Lambda}
 \end{center}
\end{figure}

With the method constructed in Sect.~\ref{sec:error}, we evaluate the uncertainty of $\tilde{X}_{\bar{K}N}$ that comes from the higher-order terms in the weak-binding relation. By adopting the Compton wavelength of the $\rho$ meson as $R_{\mathrm{typ}}$, we constrain the magnitude of the correction term $|\xi_{c}|\leq |R_{\mathrm{typ}}/R|+|l/R|^3 $. Varying complex $\xi_{c}$ with this condition in Eq.~\eqref{eq:estimate_error_quasi}, we obtain the uncertainty band of $\tilde{X}_{\bar{K}N}$ for each set as shown in  Table~\ref{tab:error_Lambda}. The results are graphically shown in Fig.~\ref{fig:error_Lambda}. In all cases, we see that the value of $\tilde{X}_{\bar{K}N}$ is larger than 1/2 even if the uncertainty band is taken into account. Thus the qualitative conclusion of the composite dominance still holds .

\begin{table}[bt]
		\caption{The results of error evaluation of the compositeness $\tilde{X}_{\bar{K}N}$ of $\Lambda (1405)$ with the value of $|R_{\mathrm{typ}}/R|$ and $|l/R|^3$}
		\label{tab:error_Lambda}
	\begin{center}
		\begin{tabular}{lccl}
			\hline
              & $|R_{\mathrm{typ}}/R|$ & $|l/R|^3 $ & $\tilde{X}_{\bar{K}N}$    \\  \hline
			 Set 1 \cite{Ikeda:2012au}  & $0.17$ & $0.14$ 
			  & $1.0^{+0.0}_{-0.4}$ \\ 
			 Set 2 \cite{Mai:2012dt}  & $0.10$ & $0.03$ 
			 & $0.6^{+0.2}_{-0.1}$  \\ 
			 Set 3 \cite{Guo:2012vv}  & $0.16$ & $0.11$ 
			 &  $0.9^{+0.1}_{-0.4}$   \\
			 Set 4 \cite{Mai:2014xna}  & $0.10$ & $0.03$ 
			 & $0.6^{+0.3}_{-0.1}$ \\ 
			 Set 5 \cite{Mai:2014xna}  & $0.12$ & $0.04$ 
			 & $0.8^{+0.2}_{-0.2}$ \\ 
			 \hline
		\end{tabular} 
	\end{center}
\end{table}

\begin{figure}[t]
 \begin{minipage}{0.5\hsize}
  \begin{center}
   \includegraphics[width=70mm]{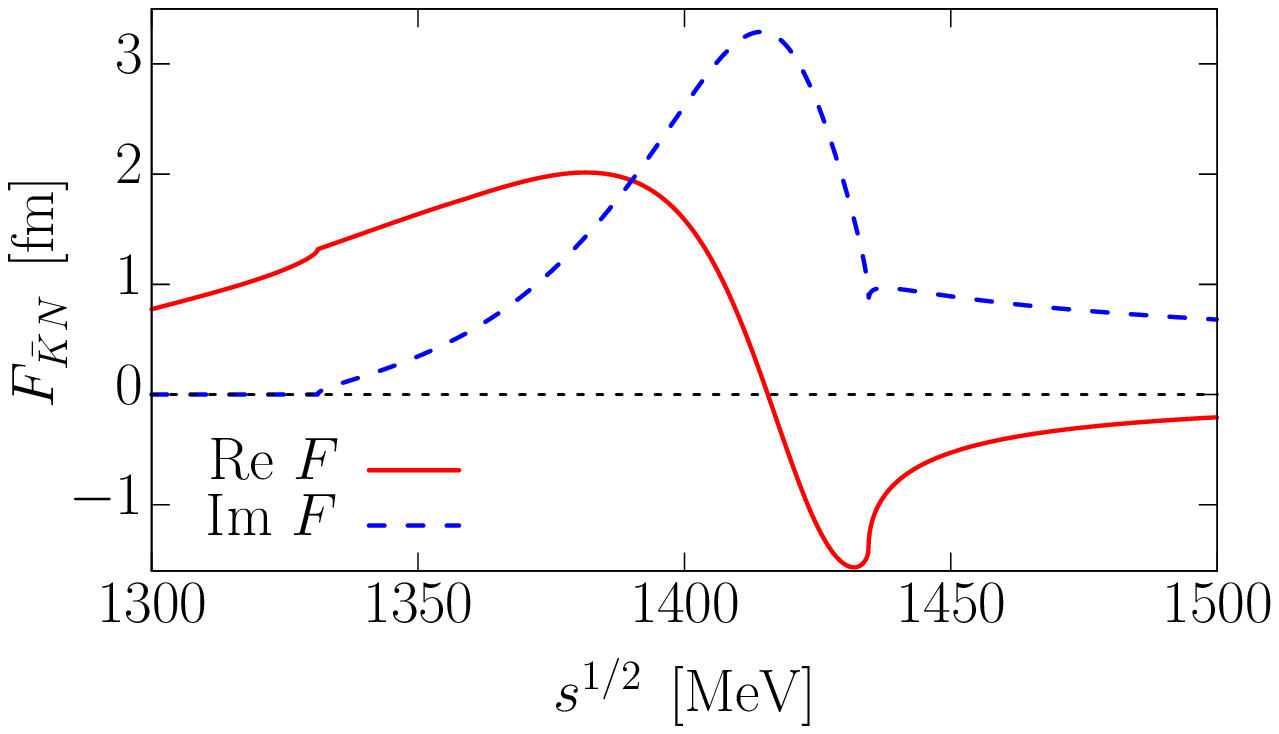}
  \end{center}
 \end{minipage}
 \begin{minipage}{0.5\hsize}
  \begin{center}
   \includegraphics[width=70mm]{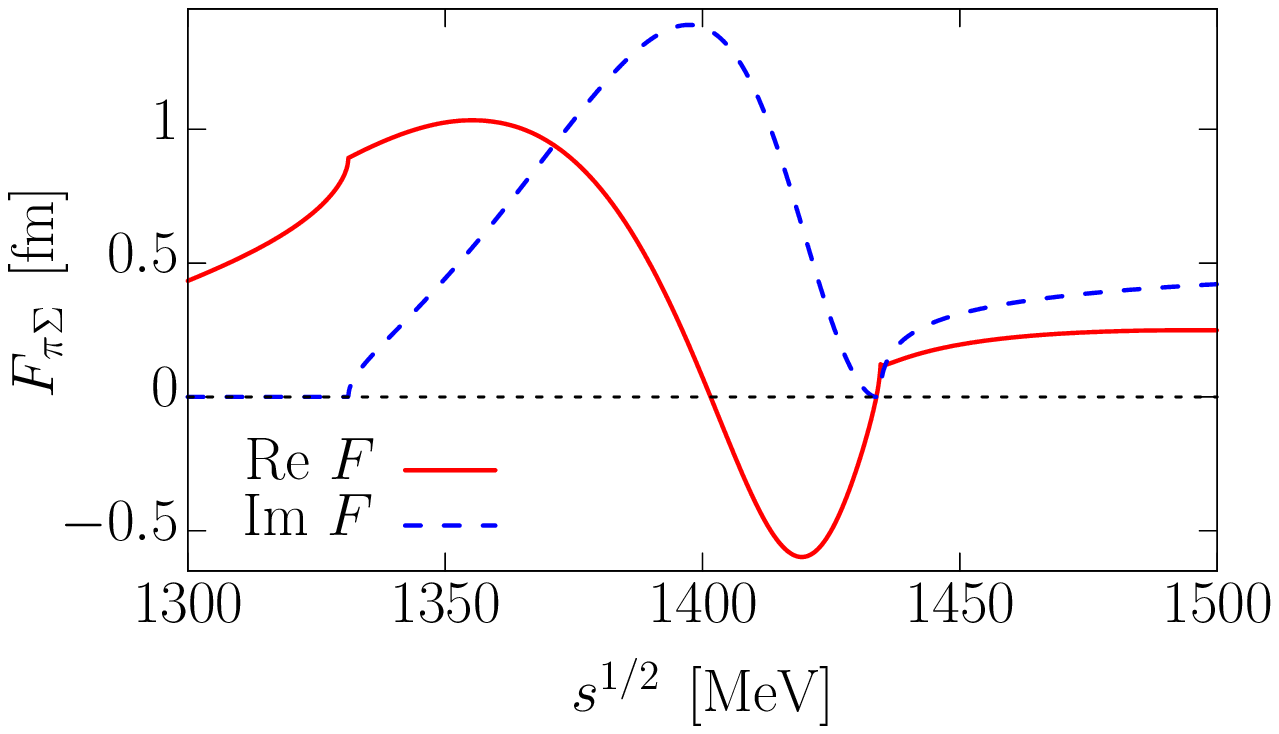}
  \end{center}
 \end{minipage}
   \caption{$I=0$ scattering amplitudes in the    $\bar{K}N\rightarrow \bar{K}N$ (right panel) and 
   $\pi\Sigma\rightarrow\pi\Sigma$ (left panel)
   channels based on Ref.~\cite{Ikeda:2012au} with the isospin-averaged hadron masses. The solid line denotes the real part and the dashed line denotes the imaginary part.}
   \label{fig:Lambda_amp}
\end{figure}

To investigate the CDD pole contribution to the $\Lambda(1405)$ state, we calculate the compositeness with extended relations \eqref{eq:comp-rel-mod-ERE-rewritten} and \eqref{eq:comp-rel-Pade-rewritten}, using the $a_0=1.39-i0.85\physdim{fm}$, $r_{\mathrm{e}}=0.24-i0.05\physdim{fm}$ and $E_h=-10-i26\physdim{MeV}$ determined from the amplitude in Ref.~\cite{Ikeda:2012au}.
Then the calculated values of $(X_{\bar{K}N}, \tilde{X}_{\bar{K}N})$ are $(1.3+i0.2, 1.0)$ with Eq.~\eqref{eq:comp-rel-mod-ERE-rewritten} and $(1.4+i0.2, 1.0)$ with Eq.~\eqref{eq:comp-rel-Pade-rewritten}.
These values are close to those of set 1 in Table~\ref{tab:Lambda}.
 This means that the ERE converges well and the CDD pole contribution can be neglected.
To check this, we show the $I=0$ scattering amplitude in the diagonal $\bar{K}N$ channel based on Ref.~\cite{Ikeda:2012au} in the left panel of Fig.~\ref{fig:Lambda_amp}. We do not find any CDD pole, which is defined as $|F(E_c)|=0$, in the energy region between $1300$ and $1500\physdim{MeV}$. This guarantees that the ERE of the $\bar{K}N$ amplitude around threshold is not disturbed by the CDD pole contribution. 
It is also instructive to look at the amplitude in the diagonal $\pi\Sigma$ channel shown in the right panel of Fig.~\ref{fig:Lambda_amp}. We notice that both the real and imaginary parts of the amplitude vanish around $1434\physdim{MeV}$. 
Namely, the $\pi\Sigma$ amplitude has a CDD pole at this energy.\footnote{In the coupled-channel scattering, each component can have a CDD pole individually. This is in contrast to the pole of the amplitude representing the eigenstate, which is determined by $\det \mathcal{F}^{-1}=0$ and the divergence appears in all the components of $\mathcal{F}_{ij}$.} 
Thus the ERE description of the $\pi\Sigma$ amplitude around its threshold will not reach the $\bar{K}N$ threshold because of the CDD pole. 
The existence of the CDD pole near the resonance pole in the $\pi\Sigma$ amplitude may be an indication of the non-$\pi\Sigma$ dominance of $\Lambda(1405)$.

In Refs.~\cite{Sekihara:2012xp,Sekihara:2014kya,Garcia-Recio:2015jsa,Guo:2015daa}, the compositeness of $\Lambda(1405)$ is also calculated in various models by evaluating the expression in Eq.~\eqref{eq:X-gG} at the pole position.
The results are summarized in Table~\ref{tab:X-papers-Lambda}.
In Refs.~\cite{Sekihara:2012xp} and \cite{Sekihara:2014kya}, the scattering amplitude is calculated from the chiral unitary approach of Refs.~\cite{Oset:2001cn} and \cite{Ikeda:2012au}, respectively. 
In the analysis of Ref.~\cite{Garcia-Recio:2015jsa}, the SU(6) model in Ref.~\cite{GarciaRecio:2005hy} is used.
In Ref.~\cite{Guo:2015daa}, the scattering amplitude based on the unitary chiral perturbation theory in Ref.~\cite{Guo:2012vv} is used.
We summarize the results in Table~\ref{tab:X-papers-Lambda}, specifying the prescription to interpret the compositeness.
We see that these studies give a consistent result of for $\bar{K}N$ dominance over the other components. 
This is also in good agreement with our model-independent results by the weak-binding relation.

In these studies, Refs.~\cite{Sekihara:2014kya} and \cite{Guo:2015daa} use the scattering amplitude in Refs.~\cite{Ikeda:2012au} and \cite{Guo:2012vv}, respectively.
Although Ref.~\cite{Guo:2015daa} uses a different prescription $|X|$ to determine the compositeness, small $U=0.1$ in set 3 indicates the difference between the prescriptions should be small, as we discussed in Sect. 3.5.
We find that the results of Refs.~\cite{Sekihara:2014kya} and \cite{Guo:2015daa} are in quantitative agreement with the corresponding results in Table~\ref{tab:Lambda}.
This agreement confirms that the magnitude of higher-order terms $\mathcal{O} (R_{\mathrm{typ}}/R)$ and $\mathcal{O}( (l/R)^3)$ are indeed small, as we have estimated.

\begin{table}
\caption{The results of the evaluation of the compositeness of $\Lambda(1405)$ at the higher-energy pole position}
\label{tab:X-papers-Lambda}
\begin{center}
\begin{tabular}{ccccc}\hline
Ref.& Amplitude & Prescription & $ X_{\bar{K}N}$ & Other components \\ \hline 
\cite{Sekihara:2012xp}& \cite{Oset:2001cn} & complex $X$ & $0.99+i0.05$ & $\phantom{-}0.01-i0.05$\\ 
\cite{Sekihara:2014kya}& \cite{Ikeda:2012au} &complex $X$ & $1.14+i0.01$ & $ -0.14-i0.01$ \\  
\cite{Garcia-Recio:2015jsa}& \cite{GarciaRecio:2005hy} &$\mathrm{Re}\hspace{1ex}X$ & $0.795$& $0.205$\\  
\cite{Guo:2015daa} & \cite{Guo:2012vv}&$|X|$& $0.81$ & $0.19$ \\\hline 
\end{tabular}
\end{center}
\end{table}

\subsection{Structure of $f_0(980)$ and $a_0(980)$}
Next we consider the scalar mesons $f_0(980)$ and $a_0(980)$, which have $I=0$ and $1$, respectively. $f_0(980)$ decays to the $\pi\pi$ channel and $a_0(980)$ decays to the $\pi\eta$ channel.
These mesons lie near the $\bar{K}K$ threshold and we study their $\bar{K}K$ compositeness.
In the following calculations, we use the isospin-averaged values for the $\bar{K}K$ threshold energy and its reduced mass.

With the analyses in Refs.~\cite{Aaltonen:2011nk,Ambrosino:2005wk,Garmash:2005rv,Ablikim:2004wn,Link:2004wx,Achasov:2000ym,Adams:2011sq,Ambrosino:2009py,Bugg:2008ig,Achasov:2000ku,Teige:1996fi}, the $\bar{K}K$ scattering amplitude is determined with the Flatt\'e parametrization~\cite{Flatte:1976xu}.
In these analyses, the heavy meson decay spectra are mainly used.
For instance,
in Ref.~\cite{Aaltonen:2011nk}, the CDF Collaboration analyses the decay process of $B_s^0\rightarrow J/\psi\pi^-\pi^+$.
The CLOE Collaboration analyses the process of $\phi\rightarrow\pi^-\pi^+\gamma$ for $f_0(980)$~\cite{Ambrosino:2005wk} and  $\phi\rightarrow\eta\pi^0\pi^0$ for $a_0(980)$~\cite{Ambrosino:2009py}.
In Ref.~\cite{Garmash:2005rv}, the Belle Collaboration studies the decay mode of $B^{\pm}\rightarrow K^\pm\pi^\pm\pi^\mp$. 
The BES Collaboration analyses the $J/\psi \rightarrow \phi\pi^+\pi^-,\phi K^+K^-$ process~\cite{Ablikim:2004wn}. 
The decay process of $D_s^+ \rightarrow \pi^+\pi^-\pi^+$ is studied by the FOCUS Collaboration in Ref.~\cite{Link:2004wx}. The processes of $e^+e^-\rightarrow\phi(1020)\rightarrow\pi^0\pi^0\gamma$ for $f_0(980)$ and $e^+e^-\rightarrow \phi(1020)\rightarrow\eta \pi^0\gamma$ for $a_0(980)$ are studied with the SND detector in Refs.~\cite{Achasov:2000ym} and \cite{Achasov:2000ku}, respectively. 
In Ref.~\cite{Adams:2011sq}, the CLEO Collaboration analyses the decay mode of $\chi_{c1}\rightarrow \eta\pi^+\pi^-$ for $a_0(980)$.
The data of the $\bar{p}p\rightarrow\eta\pi^0\pi^0$ process by the CB Collaboration are analyzed in Ref.~\cite{Bugg:2008ig} for $a_0(980)$.
In Ref.~\cite{Teige:1996fi}, the process of $\pi^-p \rightarrow \eta\pi^+\pi^-n$ is analyzed with the data of E852 experiment.

From these Flatt\'e amplitudes, the eigenenergies of $f_0(980)$ and $a_0(980)$ and the scattering lengths of the $\bar{K}K(I=0)$ and $\bar{K}K(I=1)$ channels are obtained as in Tables~\ref{tab:f0} and \ref{tab:a0}, respectively.
The small eigenenergies suggest using the weak-binding relation.\footnote{ Here we do not use the most recent analysis of $a_0(980)$ in Ref.~\cite{Adams:2011sq} because the magnitude of the eigenenergy is not small ($E_h = 31-i70\physdim{MeV}$). In this case, the weak-binding relation suffers from a substantial contribution from the higher-order terms ($|R_{\mathrm{typ}}/R|\sim0.25$ and $|l/R|^3\sim 0.13$)}
The obtained eigenenergies satisfy $|R|\gtrsim 1.5\hspace{1ex}\mathrm{fm}$ in all the analyses for both the scalar mesons.
Then we see that $|R_{\mathrm{typ}}/R| \lesssim 0.17$ and $|l/R|^3\lesssim 0.04$, 
where $R_{\mathrm{typ}}$ is determined from the $\rho$ meson exchange and 
$l =0.33 \physdim{fm}$ ($l = 0.51\physdim{fm}$) is determined from the difference between the threshold energies $\omega = 715 \physdim{MeV}$ ($\omega = 305 \physdim{MeV}$) for the $I=0$ ($I=1$) channel.
Given the estimated values, we can neglect the two higher-order terms in the weak-binding relation, and determine the $\bar{K}K$ compositeness of $f_0(980)$ and $a_0(980)$. 

\begin{table}[t]
		\caption{Properties and results for $f_{0} (980)$ quoted from Ref.~\cite{Kamiya:2015aea}: shown are the eigenenergy $E_{h}$, $\bar{K}K(I=0)$ scattering length $a_{0}$, the $\bar{K}K$ compositeness $X_{\bar{K}K}$ and $\tilde{X}_{\bar{K}K}$ and the uncertainty of the interpretation $U$}
		\label{tab:f0}
	\begin{center}
		\begin{tabular}{llllcl}
			\hline
             Ref. & $E_{h} \physdim{[MeV]}$ & $a_0 \physdim{[fm]} $ & $X_{\bar{K}K}$ & $\tilde{X}_{\bar{K}K}$ & $U/2$   \\  \hline
			 \cite{Aaltonen:2011nk}  & $\phantom{-}19-i30$ & $0.02 - i 0.95$ 
			 & $0.3-i0.3$ & $0.4$ & $0.1$   \\ 
			 \cite{Ambrosino:2005wk}  & $-\phantom{0}6-i10$ & $0.84 - i 0.85$ 
			 & $0.3-i0.1$ & $0.3$ & $0.0$  \\ 
			 \cite{Garmash:2005rv}  & $-\phantom{0}8-i28$ & $0.64 - i 0.83$ 
			 & $0.4-i0.2$ & $0.4$ & $0.0$  \\ 
			 \cite{Ablikim:2004wn}  & $\phantom{-}10-i18$ & $0.51 - i 1.58$ 
			 & $0.7-i0.3$ & $0.6$ & $0.1$  \\ 
			 \cite{Link:2004wx}  & $-10-i29$ & $0.49 - i 0.67$ 
			 & $0.3-i0.1$ & $0.3$ & $0.0$  \\ 
			 \cite{Achasov:2000ym}  & $\phantom{-}10-i\phantom{0}7$ & $0.52 - i 2.41$ 
			 & $0.9-i0.2$ & $0.9$ & $0.1$   \\ 
			 \hline
		\end{tabular} 
	\end{center}
\end{table}

\begin{table}[bt]
		\caption{Properties and results for $a_{0} (980)$ quoted from Ref.~\cite{Kamiya:2015aea}: shown are the eigenenergy $E_{h}$, $\bar{K}K(I=1)$ scattering length $a_{0}$, the $\bar{K}K$ compositeness $X_{\bar{K}K}$ and $\tilde{X}_{\bar{K}K}$ and the uncertainty of the interpretation $U$}
		\label{tab:a0}
	\begin{center}
		\begin{tabular}{llllcl}
			\hline
             Ref. & $E_{h} \physdim{[MeV]}$ & $a_0 \physdim{[fm]} $ & $X_{\bar{K}K}$ & $\tilde{X}_{\bar{K}K}$ & $U/2$  \\  \hline
			 \cite{Ambrosino:2009py} & $\phantom{0}3-i25$ & $\phantom{-}0.17 - i 0.77$ 
			 & $0.2-i0.2$ & $0.2$ & $0.0$ \\ 
			 \cite{Bugg:2008ig} & $\phantom{0}9-i36$ & $\phantom{-}0.05 - i 0.63$ 
			 & $0.2-i0.2$ & $0.2$ & $0.0$ \\ 
			 \cite{Achasov:2000ku} & $14-i\phantom{0}5$ & $-0.13 - i 2.19$ 
			 & $0.8-i0.4$ & $0.7$ & $0.1$  \\ 
			 \cite{Teige:1996fi} & $15-i29$ & $-0.13 - i 0.52$ 
			 & $0.1-i0.2$ & $0.1$ & $0.0$  \\ 
			 \hline
		\end{tabular} 
	\end{center}
\end{table}

We summarize the results for $f_0(980)$ obtained in Ref.~\cite{Kamiya:2015aea} in Table~\ref{tab:f0}. 
We see that $U$ is as small as 0.2 for all cases,
and we can consider the values of $\tilde{X}_{\bar{K}K}$ as the probabilities.
However the values of $\tilde{X}_{\bar{K}K}$ are scattered.
This is presumably caused by a large deviation among $E_h$ and $a_0$ determined by the Flatt\'e parameters.
In other words, these analyses~\cite{Aaltonen:2011nk,Ambrosino:2005wk,Garmash:2005rv,Ablikim:2004wn,Link:2004wx,Achasov:2000ym} do not give a consistent structure for $f_0(980)$.
To obtain a conclusive result, the scattering length and eigenenergy must be determined unambiguously.  
From the results in Table~\ref{tab:f0},
the uncertainty of the interpretation is expected to be small when the exact values of $a_0$ and $E_h$ are determined.
If we adopt the most recent analysis of Ref.~\cite{Aaltonen:2011nk} only, we conclude that the $\bar{K}K$ fraction of $f_0(980)$ is about one-half.

The results of $a_0(980)$ obtained in Ref.~\cite{Kamiya:2015aea} are summarized in Table~\ref{tab:a0}.\footnote{As a reference, using the analysis of  Ref.~\cite{Adams:2011sq}, where the eigenenergy is $E_h = 31-i70\physdim{MeV}$ and the scattering amplitude is $a_0 = -0.03 -i0.53\physdim{fm}$, the values of $X_{\bar{K}K}$, $\tilde{X}_{\bar{K}K}$, and $U$ are calculated as $0.2-i0.2$, $0.3$ and $0.1$, respectively.} 
Again, the small values of $U$ allow us to regard $\tilde{X}_{\bar{K}K}$ as a probability. Except for Ref.~\cite{Achasov:2000ku}, the result of $\tilde{X}_{\bar{K}K}$ is close to zero. 
Considering the experimental uncertainty of the $\bar{K}K$ coupling constant of the Flatt\'e parameters, the associated error band of $\tilde{X}_{\bar{K}K}$ is $0.7^{+0.1}_{-0.7}$ for Ref.~\cite{Achasov:2000ku}, while the magnitude of uncertainty in other analyses is smaller than $0.2$. 
Taking into account the higher-derivative terms in the weak-binding relation in Ref.~\cite{Adams:2011sq}, 
the value of $\tilde{X}_{\bar{K}K}$ is considered to be $\tilde{X}_{\bar{K}K}\lesssim0.2$. 
Thus we conclude that the structure of $a_0(980)$ is not dominated by the $\bar{K}K$ component. 
In the present analyses, we cannot pin down the physical origin of the elementariness of $a_0(980)$. 
The candidates for the dominant structure of $a_0(980)$ are, e.g., the other composite state such as $\pi\eta$, the simple $q\bar{q}$ configuration, and the tetraquark state $qq\bar{q}\bar{q}$.

\begin{table}[t]
		\caption{The results of error evaluation of the compositeness $\tilde{X}_{\bar{K}K}$ of $f_0(980)$ with the value of $|R_{\mathrm{typ}}/R|$ and $|l/R|^3$}
		\label{tab:error_f0}
	\begin{center}
		\begin{tabular}{lccl}
			\hline
             Ref. &$|R_{\mathrm{typ}}/R|$ & $|l/R|^3 $& $\tilde{X}_{\bar{K}K}$    \\  \hline
			 \cite{Aaltonen:2011nk}  & $0.17$ & $0.00$ 
			 &  $0.4^{+0.1}_{-0.2}$    \\ 
			 \cite{Ambrosino:2005wk}  & $0.01$ & $0.00$ 
			 & $0.3^{+0.1}_{-0.1}$   \\ 
			 \cite{Garmash:2005rv}  & $0.15$ & $0.01$ 
			 & $0.4^{+0.2}_{-0.1}$ \\ 
			 \cite{Ablikim:2004wn}  & $0.13$ & $0.01$ 
			 & $0.6^{+0.2}_{-0.1}$  \\ 
			 \cite{Link:2004wx}  & $0.16$ & $0.01$ 
			 & $0.3^{+0.2}_{-0.1}$   \\ 
			 \cite{Achasov:2000ym}  & $0.10$ & $0.00$ 
			 & $0.9^{+0.1}_{-0.2}$    \\ 
			 \hline
		\end{tabular} 
	\end{center}
\end{table}

\begin{table}[bt]
		\caption{The results of error evaluation of the compositeness $\tilde{X}_{\bar{K}K}$ of  $a_0(980)$ with the value of $|R_{\mathrm{typ}}/R|$ and $|l/R|^3$}
		\label{tab:error_a0}
	\begin{center}
		\begin{tabular}{lccl}
			\hline
             Ref. & $|R_{\mathrm{typ}}/R|$ & $|l/R|^3 $& $\tilde{X}_{\bar{K}K}$  \\  \hline
			 \cite{Ambrosino:2009py} & $0.14$ & $0.02$ 
			 &  $0.2^{+0.2}_{-0.1}$  \\ 
			 \cite{Bugg:2008ig} & $0.17$ & $0.04$ 
			 & $0.2^{+0.2}_{-0.1}$ \\ 
			 \cite{Achasov:2000ku} & $0.11$ & $0.01$ 
			 & $0.7^{+0.2}_{-0.1}$  \\ 
			 \cite{Teige:1996fi} & $0.16$ & $0.04$ 
			 & $0.1^{+0.2}_{-0.0}$   \\ 
			 \hline
		\end{tabular} 
	\end{center}
\end{table}

\begin{figure}
 \begin{center}
 \includegraphics[width=60mm]{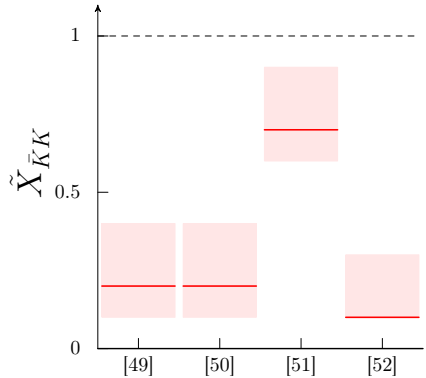}
   \caption{The results of error evaluation of the compositeness $\tilde{X}_{\bar{K}K}$ of  $a_0(980)$. The lines denote the central values and the shaded areas indicate the uncertainty bands.}
   \label{fig:error_a0}
 \end{center}
\end{figure}

We also evaluate the errors of $\tilde{X}_{\bar{K}K}$ for $f_0(980)$ and $a_0(980)$ that arise from the higher-order terms of the weak-binding relation. The results are summarized in Table~\ref{tab:error_f0} for $f_0(980)$ and in Table~\ref{tab:error_a0} and Fig.~\ref{fig:error_a0} for $a_0(980)$. We note that the experimental uncertainties of the Flatt\'e parameters are not included, which will affect the result of Ref.~\cite{Achasov:2000ku}. We see that the calculated uncertainties of $\tilde{X}_{\bar{K}K}$ are small in both cases. 
Especially for  $a_0(980)$, $\tilde{X}_{\bar{K}K} \leq 0.4$ even in the range of error, except for the result with Ref.~\cite{Achasov:2000ku}. Thus the above conclusion about the structure holds even if we take these errors into account.

In Ref.~\cite{Sekihara:2014qxa}, they calculate the compositeness of $f_0(980)$ and $a_0(980)$ using Eq.~\eqref{eq:X-gG} with the coupling constants and the eigenenergies of the Flatt\'e amplitudes in Refs.~\cite{Aaltonen:2011nk,Ambrosino:2005wk,Garmash:2005rv,Ablikim:2004wn,Link:2004wx,Achasov:2000ym,Adams:2011sq,Ambrosino:2009py,Bugg:2008ig,Achasov:2000ku,Teige:1996fi}.
The differences between the values of the compositeness come from the higher-order terms in the weak-binding relation~\eqref{eq:comp-rel-bound}, which represents the renormalization dependence in Eq.~\eqref{eq:X-gG}.
Compared with the corresponding results from the weak-binding relation in Tables~\ref{tab:f0} and \ref{tab:a0}, the absolute values of the deviations of $X_{\bar{K}K}$ are less than 0.2 in all cases, except for Ref.~\cite{Achasov:2000ku}. which has a large uncertainty in the Flatt\'e parameters. This good agreement confirms that the higher-order terms are sufficiently small.

For these scalar mesons, the compositeness is also calculated with the eigenenergy and the coupling constant in specific models obtained by phase shift analysis.
We summarize the results of $f_0(980)$ in Table~\ref{tab:X-papers-f0} and those of  $a_0(980)$ in Table~\ref{tab:X-papers-a0}.
In Ref.~\cite{Sekihara:2012xp}, they use the chiral unitary model with the Weinberg--Tomozawa interaction with dimensional regularization. 
The compositeness shown in Ref.~\cite{Sekihara:2014qxa} is calculated from the amplitude based on the chiral unitary approach with the leading-order chiral Lagrangian with a sharp cutoff. 
In Ref.~\cite{Sekihara:2014kya}, the scattering amplitude established with the coupled channel inverse amplitude method within one loop chiral perturbation theory in Ref.~\cite{GomezNicola:2001as} is employed. 
In Ref.~\cite{Guo:2015daa}, the amplitude based on the unitarized U$(3)$ chiral perturbation theory in Ref.~\cite{Guo:2012vv} is used and the absolute value of $X_{\bar{K}K}$ is interpreted as the probability.
For $f_0(980)$, the model calculations consistently show that the $\bar{K}K$ component in $f_0(980)$ is large.
If this is the case, the results with the Flatt\'e parameters in Table~\ref{tab:f0} should imply the same conclusion.
In other words, at present, the Flatt\'e analysis of the heavy meson decay does not give a consistent structure of the $f_0(980)$ meson with phase shift analysis.
In the future, it is desired to perform an analysis that takes into account the phase shift of the meson--meson scattering and the decay spectra of heavy mesons. 
For $a_0(980)$, we find only two model calculations, which consistently imply the $\bar{K}K$ fraction is not essential for the structure.
The nondominance of the$\bar{K}K$ component in $a_0(980)$ is consistent with our results in Table~\ref{tab:a0}.

In Ref.~\cite{Baru:2003qq}, the structure of these scalar mesons is also studied.
There, the authors discuss the structure of unstable states based on Weinberg's argument of the field renormalization constant using the spectral density determined with the Flatt\'e parameters.
The integrand quantity $W$ in the energy region $\pm 50 \physdim{MeV}$ around the threshold is obtained as 0.24--0.49 for $a_0(980)$ and 0.14--0.23 for $f_0(980)$. By regarding $W=(2/\pi)\arctan 2 \approx 0.70$ as the purely elementary state,
they conclude that the probability of finding $f_0(980)$ and $a_0(980)$ in a bare state are of the order of 20\% or less and about 25\% to 50\%, respectively.
This means that the $\bar{K}K$ compositeness is $X_{\bar{K}K} \gtrsim 0.8$ for $f_0(980)$ and about 0.5--0.75 for $a_0(980)$.
Although this seems to contradict our result,
the Flatt\'e parameters used in Ref.~\cite{Baru:2003qq} are different from ours. 
In fact, if our method is applied to the corresponding parameters in Ref~\cite{Baru:2003qq}, the compositeness is evaluated as 0.6--0.8 for $f_0(980)$ and 0.2--0.4 for $a_0(980)$.
Considering the uncertainties in both analyses, such as $U$, higher-order terms, the choice of the integral region of the spectral densities and the normalization of $W$, the semiquantitative conclusions are not so different.
It will be interesting to apply their method with the new Flatt\'e parameters to study scalar mesons.

\begin{table}
\caption{The results of the evaluation of  the compositeness of $f_0(980)$ at the pole position}
\label{tab:X-papers-f0}
\begin{center}
\begin{tabular}{ccccc}\hline
Ref. & Amplitude& Prescription&$ X_{\bar{K}K}$ & Other components   \\ \hline
\cite{Sekihara:2012xp}& \cite{Oller:1998hw} &complex $X$&$0.74-i0.11$ & $0.26+i0.11$\\ 
\cite{Sekihara:2014qxa}& \cite{Sekihara:2014qxa} & complex $X$&$0.70-i0.10$&$0.30+i0.10$\\
\cite{Sekihara:2014kya}& \cite{GomezNicola:2001as}& complex $X$&$0.87-i0.04$&$0.13+i0.04$\\
\cite{Guo:2015daa} & \cite{Guo:2012vv}& $|X|$&$0.65$ & $0.35$\\
\hline
\end{tabular}
\end{center}
\end{table}

\begin{table}
\caption{The results of the model evaluation of the compositeness of $a_0(980)$ at the pole position}
\label{tab:X-papers-a0}
\begin{center}
\begin{tabular}{ccccc}\hline
Ref.& Amplitude & Prescription &$ X_{\bar{K}K}$ & Other components\\ \hline
\cite{Sekihara:2012xp}&\cite{Oller:1998hw} & complex $X$&$0.38-i0.29$ &$0.62+i0.29$\\
\cite{Sekihara:2014qxa}&\cite{Sekihara:2014qxa}& complex $X$&$0.34-i0.30$& $0.66+i0.30$\\ \hline 
\end{tabular}
\end{center}
\end{table}

\section{Conclusion}

We have discussed the compositeness of unstable states around a two-body threshold in the framework of nonrelativistic effective field theory. The weak-binding relation is generalized to the unstable quasibound states, showing that the compositeness of the near-threshold unstable states is also determined from complex observables as long as the contribution from the decay mode can be neglected. The interpretation of the complex compositeness is discussed by carefully examining the condition of  the probabilistic interpretation. We have suggested a reasonable prescription to interpret complex compositeness with the real-valued compositeness $\tilde{X}$ and the uncertainty of the interpretation $U$. 
Using these new quantities, we construct a method to estimate errors that arise from the higher-order terms in the weak-binding relation.

We have presented another derivation of the weak-binding relation by separating the expansion in terms of $R_{\mathrm{typ}}/R$ from that of $R_{\mathrm{eff}}/R$.
This derivation clarifies the relation between the higher-order terms in the two different expansions and evades the implicit assumption $R_{\mathrm{eff}} \lesssim R_{\mathrm{typ}}$ in the previous derivations.
Then the weak-binding relation is improved by including the higher-order term of the effective range expansion.
We have also generalized the relation to include the contribution of the CDD pole, which is accomplished by introducing the Pad\'e approximant for the inverse amplitude.

We study the validity of the estimation of the compositeness with the weak-binding relation using models in which the exact values can be calculated.
We verify that the deviation of the estimation from the exact value is of the order of the neglected terms. 
We also see that the generalized relation including the CDD pole contribution gives a good estimation of the compositeness even for the near-threshold state with a nearby CDD pole. 
Furthermore, the method of error evaluation is examined by using solvable models with both the composite-dominant and elementary-dominant states. The results show that the exact values of the compositeness are included well within the estimated uncertainty bands, and the qualitative conclusion remains unchanged if the magnitude of the higher-order terms is small.

Finally we have applied the extended weak-binding relation for quasibound states to physical hadron resonances. 
From the threshold parameters by means of chiral SU(3) dynamics, it is concluded that $\Lambda(1405)$ has large fraction of the $\bar{K}N$ composite component.
It is also shown that the CDD pole contribution to $\Lambda(1405)$ in the $\bar{K}N$ channel is negligible by comparing the results from extended relations.
With the Flatt\'e analyses of the $\bar{K}K$ scattering, we discuss the structure of $a_0(980)$ and conclude that it has only a small $\bar{K}K$ fraction.
These conclusions are in good agreement with other studies that evaluate the compositeness at the pole position. The determination of the structure of $f_0(980)$ in the weak-binding relation is not conclusive because of the ambiguity of the Flatt\'e parameters. Comparison with the evaluation at the pole position also indicates the importance of the precise Flatt\'e parameters in the combined study with phase shift analysis.

We emphasize again that the weak-binding relations discussed in this paper
connect the internal structure of near-threshold hadron resonances with
observable quantities. Evaluation of the compositeness at the pole position,
which is a commonly adopted approach in the literature, is only possible
when a reliable theoretical amplitude is established with the help of a sufficient
amount of experimental data, and the renormalization dependence of the
compositeness is in principle unavoidable.
In contrast, the weak-binding relation can be used with a few observables (eigenenergy, scattering length, and so on), and the renormalization
dependence is suppressed thanks to the large length scale.
The applicability is however limited to near-threshold states
and the result is associated with the uncertainties by the higher-order terms.
We thus consider that the two approaches are complementary,
and both will help future investigation of the nature of hadron resonances.

\section*{Acknowledgment}
The authors thank Eulogio Oset for a valuable discussion.
This work is supported in part by JSPS KAKENHI Grant No. 24740152 and No. 16K17694 and by the Yukawa International Program for Quark-Hadron Sciences (YIPQS).



\end{document}